\documentclass[prd,superscriptaddress,preprintnumbers,tightenlines,showpacs,altaffilletter,nofootinbib,eqsecnum,amsfonts,amsmath,notitlepage,10pt]{revtex4-1}
\pdfoutput=1
\allowdisplaybreaks[1]

\usepackage{empheq}
\usepackage{psfrag,epsfig,fancyhdr}
\usepackage{graphics,avant,graphicx}
\usepackage{amsmath,amssymb,mathrsfs,amsfonts,amstext,gensymb}
\usepackage[backref=page,colorlinks=true,linkcolor=Purple,citecolor=Purple,urlcolor=Purple]{hyperref}
\usepackage{timestamp}
\usepackage[caption=false]{subfig} 
\usepackage{exscale,relsize}
\usepackage{epsf}
\usepackage{upgreek}
\usepackage{url}
\usepackage{shadethm}
\usepackage[usenames,dvipsnames,svgnames,table]{xcolor}
\usepackage[explicit]{titlesec}
\usepackage[capitalise]{cleveref}

 \DeclareMathAlphabet{\pazocal}{OMS}{zplm}{m}{n}

\usepackage{tikz}
\usetikzlibrary{arrows}
\usetikzlibrary{matrix}

\makeatletter
\newcommand*{\defeq}{\mathrel{\rlap{%
                     \raisebox{0.3ex}{$\m@th\cdot$}}%
                     \raisebox{-0.3ex}{$\m@th\cdot$}}%
                     =}
\makeatother

\newcommand{\bea}{\begin{eqnarray}}
\newcommand{\eea}{\end{eqnarray}}
\newcommand{\be}{\begin{equation}}
\newcommand{\ee}{\end{equation}}
\newcommand{\ber}{\begin{eqnarray}}
\newcommand{\eer}{\end{eqnarray}}

\def\n{\noindent}

\def\be{\begin{equation}}
\def\ee{\end{equation}}
\def\ben{\begin{eqnarray}}
\def\een{\end{eqnarray}}
\def\beqa{\begin{eqnarray}}
\def\eeqa{\end{eqnarray}}

\newcommand{\calA}{{\mathcal{A}}}
\newcommand{\calE}{{\cal E}}

\newcommand{\calH}{{\cal H}}

\newcommand{\Feyn}[1]{#1\kern-0.45em/}

\newcommand{\forget}[1]{\iffalse#1\fi}
\newcommand{\forgetmenot}[1]{\iftrue#1\fi}

\newcommand{\Sussex}{Department of Physics and Astronomy, University of Sussex, Brighton, BN1 9QH, UK}

\begin{document}


\title{Covariant Perturbations of $f(R)$ Black Holes: The Weyl Terms}

\author{Geraint Pratten}
\affiliation{\Sussex}

\timestamp
\begin{abstract}
In this paper we revisit non-spherical perturbations of the Schwarzschild black hole in the context of $f(R)$ gravity. Previous studies were able to demonstrate the stability of the $f(R)$ Schwarzschild black hole against gravitational perturbations in both the even and odd parity sectors. In particular, it was seen that the Regge-Wheeler and Zerilli equations in $f(R)$ gravity obey the same equations as their General Relativity counterparts. More recently, the 1+1+2 semi-tetrad formalism has been used to derive a set of two wave equations: one for transverse, trace-free (tensor) perturbations and one for the additional scalar modes that characterise fourth-order theories of gravitation. The master variable governing tensor perturbations was shown to be a modified Regge-Wheeler tensor obeying the same equation as in General Relativity. However, it is well known that there is a non-uniqueness in the definition of the master variable. In this paper we derive a set of two perturbation variables and their concomitant wave equations that describe gravitational perturbations in a covariant and gauge invariant manner. These variables can be related to the Newman-Penrose (NP) Weyl scalars as well as the master variables from the 2+2 formalism. As a byproduct of this study, we also derive a set of useful results relating the NP formalism to the 1+1+2 formalism valid for LRS-II spacetimes. 
\end{abstract}

\maketitle

\section{Introduction}
General Relativity (GR) has been a hugely successful theory enabling us to explain gravitational phenomena on both astrophysical and cosmological scales. The theory has passed many rigorous tests, such as solar system constraints \cite{Will14} or tests using binary pulsars \cite{Taylor82}. However, these tests only involve weak gravitational fields and/or velocities $v \ll c$, meaning that General Relativity is still essentially untested in the strong-field, highly dynamical $v \sim c$ regime, where high-energy corrections to gravitation may occur. Examples of such strong field systems include black holes and neutron stars, which are of particular interest for existing (Advanced LIGO/Virgo) \cite{Harry10} and future GW detectors. These systems are also expected to provide a testing ground in which we can understand the dynamical and phenomenological features of modified theories of gravitation in the strong field regime. In this paper, we revisit linear perturbations to the Schwarzschild black hole in both GR and $f(R)$ theories using the 1+1+2 semi-tetrad formalism. There are a few advantages to this approach. The first is that the system of equations describing the structure of the spacetime will be a set of first order differential equations in the physical curvature and dynamical variables of the covariant derivatives of tetrad vectors, unlike metric approaches which express the Einstein field equations (EFE) as a system of second order partial differential equations in terms of metric functions. Secondly, gauge-invariance will be naturally realised via the Stewart-Walker lemma \cite{Stewart74}. Finally, all objects will have a well defined physical and geometrical meaning, as the variables are related to the kinematical properties of timelike and spacelike congruences. 

Some of the most natural extensions to General Relativity are those that appear in the low energy limit of various fundamental theories. Among the most popular candidates for ultraviolet modifications to GR are the set of \textit{fourth order} theories of gravity (FOG), for which the Einstein-Hilbert action is modified by additional terms that lead to a set of field equations that are fourth order in the metric tensor. A particular subclass of fourth order theories that has received a lot of attention in the literature are the $f(R)$ theories in which the modification is some function of the Ricci scalar. A motivation for considering such a class of theories can be taken from the equivalence of metric-$f(R)$ gravity to a scalar-tensor (ST) counterpart, in which the gravitational interaction is mediated by the spin-2 graviton degrees of freedom as well as a non-minimally coupled spin-0 scalar degree of freedom. From a fundamental point of view, such ST theories arise as a natural by-product of string theory (e.g. \cite{Damour92,Damour94,Damour02}) and date back to the original work by Jordan \cite{Jordan59}, Fierz \cite{Fierz56} and Brans and Dicke \cite{Brans61}. 

In order for $f(R)$ theories to be viable alternatives to General Relativity, there are a number of minimal constraints that we can impose. For example, they must reproduce cosmological dynamics consistent with observations, they must be free from tachyonic instabilities and ghosts and they must reproduce acceptable Newtonian and post-Newtonian limits. In addition, we can also demand that certain well defined solutions in General Relativity, such as the Schwarzschild solution, be stable against perturbations in any theory of modified gravity. The analysis of linear non-spherical perturbations of a Schwarzschild black hole in ST theories has previously been considered in \cite{DeFelice11,Myung11,Nzioki14}.

In General Relativity, the Jebsen-Birkhoff theorem constrains spherically symmetric vacuum spacetimes to be either static or spatially homogeneous. It has been shown that the rigidity of the theorem is upheld even upon the introduction of perturbations. Notably, almost spherical symmetry and/or almost vacuum imply an almost static or almost spatially homogeneous spacetime \cite{Goswami11,Goswami12,Ellis13}. However, it is not clear that this theorem should necessarily hold for theories with additional degrees of freedom, as is the case for scalar-tensor and fourth order theories. Importantly, \cite{Nzioki13} found a non-zero measure in the parameter space of $f(R)$ theories for which a Jebsen-Birkhoff like theorem exists. This provides us with an additional set of constraints that guarantee the stability of a Schwarzschild solution under generic perturbations. Using these results, \cite{Nzioki14} applied the 1+1+2 semi-tetrad formalism to study non-spherical linear perturbations to the Schwarzschild black hole in $f(R)$ gravity. Working in the Jordan frame, \cite{Nzioki14} derived a modified Regge-Wheeler tensor that unifies the axial and polar degrees of freedom into a single transverse-traceless tensor obeying a tensorial form of the Regge-Wheeler equation \cite{Clarkson03,Pratten14}. 

In this paper, however, we advocate that a cleaner way to study the evolution of gravitational perturbations in the Schwarzschild spacetime is via the Weyl tensor. In particular, we introduce a perturbation variable $\pazocal{J}_{ab}$ constructed from the electric and magnetic Weyl 2-tensors. We show that this variable obeys a covariant, gauge-invariant and frame-invariant wave equation that is valid in both the polar and axial sectors. As such, this variable unifies the Regge-Wheeler and Zerilli equations into a single compact form. Unlike the modified Regge-Wheeler variable for $f(R)$ gravity, this perturbation variable demonstrates the decoupling of the scalar and gravitational wave modes in a clean and transparent manner. All equations presented here will be valid in both $f(R)$ and General Relativity unless explicitly stated otherwise. Secondly, we introduce a purely axial variable $\pazocal{V}_{ab}$ that constitutes a covariant, gauge-invariant generalisation of the Regge-Wheeler (RW) variable. This variable can also be used to re-interpret the RW term as the radial part of the magnetic Weyl tensor. The variable $\pazocal{V}_{ab}$ also obeys a covariant and gauge-invariant wave equation that reduces to the RW equation when restricting ourselves to the RW tortoise coordinates. 

Finally, we argue that the 1+1+2 formalism provides a powerful tool for understanding the physical and geometrical interpretation of other approaches to perturbation theory in General Relativity. Notably, we see that the Newman-Penrose (NP) Weyl scalars $\Psi_{0}$ and $\Psi_4$ can be related to our perturbation variable $\pazocal{J}_{ab}$ and the imaginary part of the NP Weyl scalar $\Psi_2$ can be related to our perturbation variable $\pazocal{V}_{ab}$. Likewise, for the 2+2 formalism, the polar master variables $\chi$ and $\varphi$ are just related to the electric Weyl 2-tensor $\calE_{ab}$ and the master variable $\varsigma$ is related to $\mathcal{H}_{ab}$. In the axial sector, the master variable $\Pi$ can be exactly related to $\pazocal{V}_{ab}$, as is expected. A roadmap to the set of master equations in the 2+2 formalism for $f(R)$ gravity is highlighted based on the results presented here. 

This paper is organised as follows. In Sec. \ref{sec:MG} we summarise fourth order theories of gravity and summarise the approach taken to model the $f(R)$ theory. In Sec. \ref{sec:1p3} we introduce the 1+3 formalism as the starting point for the 1+1+2 approach used here. Sec. \ref{sec:1+1+2} builds on the previous Section and introduces the 1+1+2 formalism and defines all necessary variables and operators that will be required in subsequent sections. In Sec. \ref{sec:WT} we start with review of the previous approaches to studying gravitational perturbations in the Schwarzschild spacetime with the 1+1+2 approach including the extension to $f(R)$. We then introduce a set of new perturbation variables derived from the Weyl tensor that are central to the remainder of the paper. In Sec. \ref{sec:OtherFormalism} we use the results from the previous section to provide intuition and insight into other approaches to perturbations of the Schwarzschild spacetime. Notably, we discuss the geometric interpretation of the NP Weyl and NP Ricci scalars before discussing the physical interpretation of the 2+2 gauge invariant master variables. 

\section{Modified Gravity and Black Holes}
\label{sec:MG}
\subsection{Fourth Order Gravity}
General Relativity is a unique four-dimensional theory for which the gravitational interactions are mediated by a massless spin-2 particle, the graviton, and the field equations are second order. In fourth order gravity, we consider the following modification to the Einstein-Hilbert action
\begin{align}
\int d^4 x \, \sqrt{-g} \, R \rightarrow \int d^4 x \, \sqrt{-g} \, f \left( R, R_{ab} R^{ab} , C_{abcd} C^{abcd} \right) .
\end{align}
\n
As the curvature functions contain second derivatives of the metric, the resulting field equations are fourth order. In fact, a consequence of Lovelock's theorem is that the field equations for a metric theory of modified gravity in a four-dimensional Riemannian manifold will admit higher than second order derivatives \cite{Lovelock71,Lovelock72}. Such higher order derivatives are potentially problematic as they typically lead to Ostrogradski instabilities \cite{Ostrogradski}. The sub-class of $f(R)$ theories evade these instabilities as they are degenerate, i.e. the highest derivative term cannot be written as a function of canonical variables. This can be seen through the fact that in the Ricci scalar $R$, only a single component of the metric appears with second derivatives. The concomitant new degree of freedom can then be completely fixed by the $g_{00}$ constraint preventing ghost instabilities from arising in $f(R)$ theories. Alternatively, the $f(R)$ class of theories can simply be re-cast as a scalar-tensor theory for which the gravitational interaction is mediated by a spin-0 scalar as well as the spin-2 graviton degrees of freedom \cite{Sotiriou08}. Were we to consider theories involving curvature invariants $R_{a b} R^{ab}$, $R_{a b c d} R^{a b c d}$ or Weyl invariants $C_{abcd} C^{abcd}$, these would be non-degenerate and our resulting theory would be plagued by Ostrogradski instabilities. In this paper, we restrict ourselves to the Ostrogradski stable $f(R)$ theories. 

\subsection{$f(R)$ Gravity}
The action for $f(R)$ gravity can be written as a simple generalisation of the Einstein-Hilbert action
\begin{align}
S &= \frac{1}{2} \int \, dV \, \left[ \sqrt{-g} \, f(R) + 2 \mathcal{L}_M (g_{ab} , \psi ) \right] ,
\end{align}
\n
where $\mathcal{L}_M$ corresponds to matter fields present in the theory. Upon variation with respect to the metric, the perturbed action can be written as
\begin{align}
\delta S &= - \frac{1}{2} \int dV \, \sqrt{-g} \, \left[ \frac{1}{2} f g_{ab} \delta g^{ab} - f^{\prime} \delta R + T^M_{ab} \, \delta g^{ab} \right] 
\end{align}
\n
where, following the notation of \cite{Nzioki14}, a $^{\prime}$ denotes differentiation with respect to $R$ and $T^M_{ab}$ is the standard energy-momentum tensor of matter. Demanding that the action be stationary with respect to variations in the metric, it follows that
\begin{align}
f^{\prime} \left( R_{ab} - \frac{1}{2} g_{ab} R \right) &= \frac{1}{2} g_{ab} \, \left( f - R f^{\prime} \right) + \nabla_a \nabla_b f^{\prime} - g_{ab} \Box f^{\prime} + T^M_{ab} .
\end{align}
\n
Clearly, if $f=R$ then the field equations reduce to those of General Relativity. In the covariant approach to fourth order gravity, the field equations for $f(R)$ can be re-expressed in the form of effective Einstein field equations
\begin{align}
G_{ab} &= \left( R_{ab} - \frac{1}{2} g_{ab} R \right) = T_{ab} 
\end{align}
\n
where $T_{ab}$ is now a combination of two effective fluids corresponding to an effective curvature fluid and an effective matter fluid,
\begin{align}
T_{ab}& = \tilde{T}^M_{ab} + T^R_{ab} \\
\tilde{T}^M_{aB} &= \frac{T^M_{ab}}{f^{\prime}} \\
T^R_{ab} &= \frac{1}{f^{\prime}} \, \left[ \frac{1}{2} g_{ab} \left(f - R f^{\prime} \right) + \nabla_a \nabla_b f^{\prime} - g_{ab} \Box f^{\prime} \right]
\label{eqn:EM_Curvature}
\end{align}
\n
The key advantage to this representation is that it is much easier to adapt techniques from the covariant approaches to relativistic cosmology to $f(R)$ gravity \cite{Ellis99,Nzioki14}. 

In the field equations presented above, we have terms involving higher than second order derivatives of the metric. For example, in \cref{eqn:EM_Curvature}, we have terms of the form $\nabla_a \nabla_b f^{\prime}$. In non-degenerate theories, such terms would be problematic. 

As per \cite{Nzioki14}, the calculations are restricted to the Jordan frame for which matter is minimally coupled and the gravitational scalar is non-minimally coupled to the curvature. In this frame, the dynamics of the extra gravitational degree of freedom will be determined by trace of the effective Einstein field equations (EFE) resulting in a linearised scalar wave equation for the Ricci scalar. 

\subsection{Schwarzschild Black Hole in f(R)}
In \cite{Nzioki10,Nzioki13}, it was demonstrated that for $f(R)$ theories, where $f$ is of class $C^3$ at $R=0$ with the conditions $f(0) = 0$ and $f^{\prime} (0) \neq 0$, any almost spherically symmetric solution with almost vanishing Ricci scalar in empty space for some open set $\mathcal{S}$ will be locally almost equivalent a part of the maximally extended Schwarzschild solution in $\mathcal{S}$. This important result constitutes a Jebsen-Birkhoff like theorem that details the conditions required for the existence of Schwarzschild spacetimes in $f(R)$ theories. 

As $f$ is of class $C^3$ at $R=0$, it is implied that \cite{Nzioki13}
\begin{align}
| f^{\prime} (0) | < + \infty , \qquad | f^{\prime \prime} (0) | < + \infty , \qquad | f^{\prime \prime \prime} (0) | < + \infty ,
\end{align}
\n
with the constraint that \cite{Nzioki13}
\begin{align}
f(0) = 0  \qquad R = 0. 
\end{align}
\n
Performing a Taylor series expansion about the background, the linear order result yields 
\begin{align}
\label{eqn:fRSch}
f(R) &= f^{\prime}_0 R .
\end{align}

\section{The 1+3 Covariant Formalism}
\label{sec:1p3}
\subsection{The Formalism}
\label{sec:1+3}
We use a covariant and gauge-invariant approach built on a local 1+3 threading of spacetime with respect to a preferred timelike congruence \cite{Ellis99,Tsagas07}. Consider a spacetime $( \mathcal{M} , g)$ and introduce a preferred timelike congruence $u^a$ generated by a set of observers with 4-velocity 
\begin{align}
u^a = \frac{d x^a}{d \tau} , \qquad u^a u_a = -1,
\end{align}
\n 
where $\tau$ is the proper time as measured by the fundamental observers. The spacetime is then locally split in the form $R \otimes H$ with $R$ denoting the timeline along $u^a$ and $H$ denoting the tangent 3-spaces perpendicular to $u^a$. The induced metric tensor on the 3-spaces is $h_{ab} = g_{ab} + u_a u_b$ and can be used to project tensors onto the spatial surfaces.  For example, any 4-vector $X^a$ can be decomposed into a 3-vector component defined on the 3-space and a (1+3) scalar component parallel to the congruence:
\begin{align}
X^a &= \pazocal{X}^a_{\perp} + u^a \pazocal{X}_{\parallel} \qquad \textrm{where} \qquad \pazocal{X}^a_{\perp} = h^a_{\phantom{a} b} X^b \quad \textrm{and} \quad \pazocal{X}_{\parallel} = u_b X^b 
\end{align}
\n
Similarly, any second rank tensor may be irreducibly split into scalar, 3-vector and \textit{projected, symmetric, trace-free (PSTF)} 3-tensor components:
\begin{align}
X_{ab} &=  \underbrace{ \frac{1}{3} \pazocal{X} \, h_{ab} }_{\textcolor{violet}{\text{scalar}}} \; + \; \overbrace{ \pazocal{X}_{[a b]} }^{\textcolor{violet}{\text{3-vector}}}  \; + \;   \underbrace{ \pazocal{X}_{\langle a b \rangle}  }_{\textcolor{violet}{\text{3-tensor}}}  , 
\end{align}
\n
where the 3-vector component can be re-expressed in terms of a genuine 3-vector and the alternating tensor $\epsilon_{abc} = u^d \, \eta_{dabc}$. Together, the timelike congruence $u^a$ and induced metric $h_{ab}$ allow us to define two preferred derivative operators. The first operator is a convective derivative along $u^a$
\begin{align}
\dot{\pazocal{X}}_{a \dots b}^{\phantom{a \dots b} c \dots d} &\defeq u^e \, \nabla_e \, \pazocal{X}_{a \dots b}^{\phantom{a \dots b} c \dots d}
\end{align}
\n
and the second is a totally projected spatial derivative defined in the 3-surfaces orthogonal to $u^a$
\begin{align}
 D_e \, \pazocal{X}^{a \dots b}_{\phantom{a \dots b} c \dots d} &\defeq h^a_{\phantom{a} m} \dots h^b_{\phantom{b} n} h_c^{\phantom{c} p} \dots h_d^{\phantom{d} q} h^r_{\phantom{r} e} \, \nabla_r \pazocal{X}^{m \dots n}_{\phantom{p \dots q} p \dots q}  .
\end{align}

\subsection{The Variables}
The three main groups of variables in the 1+3 formalism are the set of kinematical variables describing the flow of $u^a$, the set of variables associated with the gravitational field as well as the variables describing the energy-momentum tensor. The kinematical variables arise from the irreducible decomposition of the covariant derivative of the timelike congruence:
\begin{align}
\nabla_b u_a &= \dot{u}_a u_b + \frac{1}{3} \Theta h_{ab} + \sigma_{ab} + \omega_{ab} ,
\end{align}
\n
where $\dot{u}_a$ is the acceleration of the congruence, $\sigma_{ab}$ the volume preserving shear, $\omega_{ab}$ the vorticity and $\theta$ the expansion of the congruence. These variables describe the kinematical properties of the flow associated to the observer $u^a$. 

The gravitational field is covariantly described by the Ricci tensor and the Weyl tensor. The Ricci tensor describes the local gravitational field at a given point due to the local matter content, as can be seen via the EFE. The Weyl tensor describes the non-local gravitational field that is mediated by gravitational waves and tidal forces. These two objects allow us to decompose the gravitational field into the local and non-local terms as follows:
\begin{align}
R_{abcd} &= C_{abcd} + \frac{1}{2} \left( g_{ac} R_{bd} + g_{bd} R_{ac} - g_{bc} R_{ad} - g_{ad} R_{bc} \right) - \frac{R}{6} \left( g_{ac} g_{bd} - g_{ad} g_{bc} \right) .
\end{align}
\n
In addition, we can decompose the Weyl tensor into its irreducible parts 
\begin{align}
E_{ab} &= C_{acbd} u^c u^d \qquad \textrm{and} \qquad H_{ab} = \frac{1}{2} \epsilon_{a}^{\phantom{a} cd} C_{cdbe} u^e ,
\end{align}
\n
where $E_{ab}$ denotes the electric Weyl tensor and $H_{ab}$ the magnetic Weyl tensor. The electric Weyl tensor generalises the Newtonian tidal tensor whereas the magnetic Weyl tensor has no Newtonian counterpart as it describes genuine relativistic effects such as frame dragging. 

\subsection{The Equations}
In the 1+3 approach, the evolution and constraint equations for the variables arises from two sets of identities satisfied by the Riemann tensor with the EFE only being used to algebraically substitute the Ricci tensor for the equivalent energy-momentum terms. The first set of equations are given by the Ricci identities applied to $u^a$
\begin{align}
2 \nabla_{[a} \nabla_{b]} u_c &= R_{abcd} u^d ,
\end{align}
\n
a second set of equations are provided by the once contracted Bianchi identities,
\begin{align}
\nabla^d C_{abcd} &= \nabla_{[b} R_{a]c} + \frac{1}{6} g_{c[b} \nabla_{a]} R ,
\end{align}
\n
and the last set of equations are provided by the twice contracted Bianchi identities 
\begin{align}
\nabla^b T_{ab} = 0 .
\end{align}
\n
The algebraic relations for the Ricci tensor are given by EFE
\begin{align}
R_{ab} &= T_{ab} - \frac{1}{2} T g_{ab} + \Lambda g_{ab} .
\end{align}

\section{The 1+1+2 Covariant Formalism}
\label{sec:1+1+2}

\subsection{The Formalism}
The 1+3 formalism has been hugely successful in relativistic cosmology as it is particularly adapted to the models that it aims to describe. For homogeneous and isotropic cosmologies, the 1+3 formalism covariantly factorises out the only essential coordinate, time. The background spacetime is described by a system of ODEs for a set of 1+3 covariant scalar variables and all conventional ODE methods may be invoked. As the 3-vectors and 3-tensors vanish in the background due to the symmetry, they are implicitly gauge invariant as per the Stewart-Walker lemma \cite{Stewart74}. Covariant and gauge-invariant linear is realised in this approach by noting that the non-vanishing background variables will be zeroth order and all variables that vanish in the background will be first order. We can then linearise the full system of equations about a specified background, dropping all terms that are second order and higher. Under this procedure, all vector-tensor and tensor-tensor couplings are killed off as they are implicitly second order or higher. The result is a system of equations that is tractable and may be solved.

If we relax the symmetry of the spacetime, there will be non-zero vectors and tensors in the background and the 1+3 system of equations is rendered intractable. However, if the spacetime possesses a preferred spatial direction at each point (such as spherical symmetry, local rotational symmetry or $G_2$ spacetimes) we can introduce an additional frame vector such that it covariantly factors out the two essential coordinates: time and the preferred radial direction. Consequently, all vectors and tensors will vanish on the 2-surfaces of homogeneity resulting in a system of ODEs for 1+1+2 scalar variables. Historically, such decompositions were originally introduced by \cite{Greenberg70} and were further developed by \cite{Tsamparlis83,Mason85,Tsamparlis92,Clarkson03,Betschart04,Burston06,Clarkson07}. In this paper, we follow the approach to the 1+1+2 formalism as given in \cite{Clarkson03,Clarkson07}. 

In the 1+1+2 formalism we start by introducing a preferred spacelike congruence $n^a$ that is orthogonal to $u^a$ such that
\begin{align}
n_a u^a = 0, \qquad n_a n^a = 1 .
\end{align}
\n
The induced metric on the tangent 2-surfaces orthogonal to $n^a$ and $u^a$  is defined by
\begin{align}
N_{ab} &= h_{ab} - n_a n_b = g_{ab} + u_a u_b - n_a n_b ,
\end{align}
\n
and can be used to project all vectors and tensors orthogonal to $u^a$: $n^a N_{ab} = u^a N_{ab} = 0$. Any 3-vector $\psi_a$ may be irreducibly split into a scalar ${\bf{\Psi}}$ parallel to $n^a$ and a 2-vector ${\bf{\Psi}}_a$ lying in the 2-surfaces orthogonal to $n^a$:
\begin{align}
\psi_a &= {\bf{\Psi}} n_a + {\bf{\Psi}}_a .
\end{align}
\n
Similarly, any 3-tensor $\psi_{ab}$ can be irreducibly split into a scalar, 2-vector and PSTF (with respect to $n^a$) 2-tensor:
\begin{align}
\psi_{ab} &=  \underbrace{ {\bf{\Psi}} }_{\textcolor{violet}{\text{scalar}}} \left( n_a n_b - \frac{1}{2} N_{ab} \right) \; + \; \overbrace{ 2 {\bf{\Psi}}_{(a} n_{b)} }^{\textcolor{violet}{\text{2-vector}}} \; + \; \underbrace{ {\bf{\Psi}}_{\lbrace ab \rbrace} }_{\textcolor{violet}{\text{2-tensor}}} ,
\end{align}
\n
where 
\begin{align}
{\bf{\Psi}} &= n^a n^b \psi_{ab} , \qquad {\bf{\Psi}}_a = N_a^{\phantom{a} b} n^c \psi_{bc} \qquad \textrm{and}  \qquad {\bf{\Psi}}_{\lbrace a b \rbrace} = \psi_{\lbrace a b \rbrace} = \left( N_{(a}^{\phantom{(a} c} N_{b)}^{\phantom{b)} d} - \frac{1}{2} N_{ab} N^{cd} \right) \psi_{cd} .
\end{align}
\n 
In the above, we have introduced the notation ${\bf{\Psi}}_{\lbrace a \dots c \rbrace}$ to denote PSTF with respect to $n^a$. As in the 1+3 formalism, the existence of the preferred spatial congruence and induced metric $N_{ab}$ allows us to define two derivative operators. The first is a convective derivative along the spacelike congruence
\begin{align}
\hat{\psi}_{a \dots b}^{\phantom{a \dots b} c \dots d} &\defeq n^e \, D_e \, \psi_{a \dots b}^{\phantom{a \dots b} c \dots d} ,
\end{align}
\n
and the second is a totally projected derivative on the 2-surfaces 
\begin{align}
 \delta_e \psi_{a \dots b}^{\phantom{a \dots b} c \dots d} &\defeq N_e^{\phantom{e} j} \, N_a^{\phantom{e} f} \dots N_b^{\phantom{e} g} N_h^{\phantom{e} c} \dots N_i^{\phantom{e} d} \, D_j \, \psi_{f \dots g}^{\phantom{a \dots b} h \dots i} .
\end{align}
We can now decompose the covariant spatial derivative of $n^a$ into its irreducible parts\cite{Clarkson03,Clarkson07}
\begin{align}
D_a n_b &= n_a a_b + \frac{1}{2} \phi N_{ab} + \xi \epsilon_{ab} + \zeta_{ab} ,
\end{align}
\n
where 
\begin{align}
a_a &= n^c D_c n_a = \hat{n}_a ,\\
\phi &= \delta_a n^a ,\\
\xi &= \frac{1}{2} \epsilon^{ab} \delta_a n_b ,\\
\zeta_{ab} &= \delta_{\lbrace a} n_{b \rbrace} .
\end{align}
\n
Here, we interpret $\phi$ as the expansion of the 2-surface, $\zeta_{ab}$ as the distortion of the 2-surface (or shear of $n^a$), $a^a$ the acceleration and $\xi$ the rotation of $n^a$. Likewise, we can also perform an irreducible split of the timelike derivative of $n^a$,
\begin{align}
\dot{n}_a &= \mathcal{A} u_a + \alpha_a \qquad \textrm{where} \quad \alpha_a = \dot{n}_a \quad \textrm{and} \quad \mathcal{A} = n^a \dot{u}_a .
\end{align}
The remaining 1+1+2 variables arise from the irreducible decomposition of the usual kinematical and gravitational variables in the 1+3 formalism \cite{Clarkson03,Clarkson07}:
\begin{align}
\dot{u}^a &= \calA n^A + \calA^a \\
\omega^a &= \Omega n^a + \Omega^a \\
\sigma_{ab} &= \Sigma \left( n_a n_b - \frac{1}{2} N_{ab} \right) + 2 \Sigma_{(a} n_{b)} + \Sigma_{ab} \\
E_{ab} &= \calE \left( n_a n_b - \frac{1}{2} N_{ab} \right) + 2 \calE_{(a} n_{b)} + \calE_{ab} \\
H_{ab} &= \calH \left( n_a n_b - \frac{1}{2} N_{ab} \right) + 2 \calH_{(a} n_{b)} + \calH_{ab} .
\end{align}
The key variables in the 1+1+2 formalism for vacuum spacetimes are therefore
\begin{align}
\lbrace \Theta, \calA, \phi, \Sigma, \calE, \Omega, \calH, \xi, \calA^a, \Omega^a, \Sigma^a, \alpha^a, a^a, \calE^a, \calH^a, \Sigma_{ab}, \calE_{ab}, \calH_{ab}, \zeta_{ab} \rbrace
\end{align}
\n
In the extension to studying perturbations to the $f(R)$ Schwarzschild black hole, we must also implicitly include the 1+1+2 energy-momentum variables in order to describe the effective curvature fluid. The 1+1+2 energy-momentum tensor can be written as follows:
\begin{align}
T_{ab} &= \mu u_a u_b + p \left( N_{ab} - n_a n_b \right) + 2 Q u_{(a} n_{b)} + 2 u_{(a} Q_{b)} + \Pi \left( n_a n_b - \frac{1}{2} N_{ab} \right) + \Pi_{(a} n_{b)} + \Pi_{ab} .
\end{align}
\n
The full decomposition of the curvature fluid energy-momentum tensor into its 1+1+2 form is given by \cite{Nzioki14}. The linearisation procedure around a Schwarzschild proceeds by substituting \cref{eqn:fRSch} into \cref{eqn:EM_Curvature} and performing a 1+1+2 decomposition. Zeroth order terms are given by the set of Schwarzschild scalars $\lbrace \calE, \phi, \calA \rbrace$ and all products of first order terms or higher are dropped from the resulting expressions, leading to the following definitions \cite{Nzioki14}
\begin{align}
\mu^R &= \frac{1}{f^{\prime}_0} \left[ f^{\prime \prime}_0 \hat{\hat{R}} + \phi f^{\prime \prime}_0 \hat{R} + f^{\prime \prime}_0 \delta^2 R \right] , \\
p^R &= \frac{1}{f^{\prime}_0} \left[ f^{\prime \prime}_0 \ddot{R} - \mathcal{A} f^{\prime \prime}_0 \hat{R} - \frac{2}{3} \phi f^{\prime \prime}_0 \hat{R} - \frac{2}{3} f^{\prime \prime}_0 \delta^2 R - \frac{2}{3} f^{\prime \prime}_0 \hat{\hat{R}} \right], \\
Q^R &= - \frac{1}{f^{\prime}_0} \left[ f^{\prime \prime}_0 \left( \dot{\hat{R}} - \mathcal{A} \dot{R} \right) \right] , \\
Q^R_a &= - \frac{f^{\prime \prime}_0}{f^{\prime}_0} \delta_a \dot{R} , \\
\Pi^R &= \frac{1}{f^{\prime}_0} \left[ \frac{2}{3} f^{\prime \prime}_0 \hat{\hat{R}} - \frac{1}{3} \phi f^{\prime \prime}_0 \hat{R} - \frac{1}{3} f^{\prime \prime}_0 \delta^2 R \right] , \\
\Pi^R_a &= \frac{1}{f^{\prime}_0} \left[ f^{\prime \prime}_0 \delta_a \hat{R} - \frac{1}{2} \phi f^{\prime \prime}_0 \delta_a R \right] , \\
\Pi^R_{ab} &= \frac{f^{\prime \prime}_0}{f^{\prime}_0} \, \delta_{\lbrace a} \delta_{b \rbrace} R .
\end{align}
\n
The identities
\begin{align}
\dot{\Pi}^R_{\lbrace ab \rbrace} &= - \delta_{\lbrace a} Q_{b \rbrace} \qquad \textrm{and} \qquad \epsilon_{c \lbrace a} \hat{\Pi}_{b \rbrace}^{\phantom{b \rbrace} c} - \epsilon_{c \lbrace a} \delta^c \Pi_{b \rbrace} + \frac{1}{2} \phi \epsilon_{c \lbrace a} \Pi_{b \rbrace}^{\phantom{b \rbrace} c} = 0 , 
\end{align}
\n
have been used to simplify the evolution and propagation equations for the electric and magnetic Weyl 2-tensors as given in \cref{eqn:WeylPropEv}.

\subsection{The Equations}
The full system of 1+1+2 equations arises from the decomposition of the 1+3 equations coupled with a further system of equations defined by the Ricci identities for $n^a$.
\begin{align}
R_{abc} &= 2 \nabla_{[a} \nabla_{b]} n_c - R_{abcd} n^d = 0 .
\end{align}
\n
The Ricci identity applied to $n^a$ guarantees that we have sufficient equations in order to determine the $1+1+2$ variables. In particular, this third rank tensor can be covariantly split with respect to $u^a$ and $n^a$ to give dynamical equations for the variables derived from the irreducible decomposition of $n^a$ $(\alpha_a , a_a , \phi , \xi , \zeta_{ab})$ \cite{Clarkson03,Clarkson07}. The complete system of equations can be covariantly split into propagation, evolution and constraint equations by projections with respect to $u^a, n^a, N^{ab}$ as well as various symmetry operations. Evolution equations involve dot derivatives, propagation equations involve hat derivatives, mixed propagation and evolution involve both and constraint equations are restricted to the 2-surfaces. The full system of 1+1+2 equations can be found in \cite{Clarkson03,Clarkson07} and the system of 1+1+2 equations for the $f(R)$ Schwarzschild spacetime in \cite{Nzioki14}. We will not reproduce these equations here.  

\section{The Regge-Wheeler Tensor}
\label{sec:WT}
\subsection{Review: The Background Spacetime}
The Schwarzschild spacetime is covariantly characterised by the following non-zero 1+1+2 scalars: $\lbrace \calA , \calE, \phi \rbrace$. The system of equations that covariantly describes the Schwarzschild spacetime can be written in the following compact form:
\begin{align}
\hat{\phi} &= - \frac{1}{2} \phi^2 - \calE, \\
\hat{\calE} &= - \frac{3}{2} \calE \phi , \\
\hat{\calA} &= - \calA \left( \phi + \calA \right) ,
\end{align}
\n
with the constraint equation
\begin{align}
\calE + \phi \calA &= 0 .
\end{align}
\n
A parametric solution for these equations can be found \cite{Clarkson03}:
\begin{align}
\phi &= \frac{2}{r} \sqrt{1 - \frac{2 M}{r}} , \qquad \calA = \frac{M}{r^2} \left[ 1 - \frac{2 M}{r} \right]^{-\frac{1}{2}} , \qquad \calE = - \frac{2 M}{r^3} .
\label{eqn:bgpara}
\end{align}

\subsection{Review: Gauge Invariant Perturbations}
In more conventional approaches to linear perturbation theory in GR, the perturbation to some tensor quantity is the difference between its value at some event in the physical spacetime and its value at the corresponding event in the background spacetime. Following \cite{StewartBook}, if we consider a background 4-dimensional spacetime $\mathcal{M}_0$ then we can introduce a 5-dimensional smooth manifold that contains a 1-parameter family of smooth non-intersecting 4-manifolds $\mathcal{M}_{\epsilon}$. These manifolds are interpreted as perturbations to the background spacetime. Implicitly, this means that to define a perturbation we must choose a gauge, or a mapping from $\mathcal{M}_0$ to $\mathcal{M}_{\epsilon}$. If $Q$ is some geometric object in the background and $Q_{\epsilon}$ the corresponding object in the perturbed spacetime then we can define a Taylor series expansion along the integral curves of a vector field $V$ that is everywhere orthogonal to $\mathcal{M}_{\epsilon}$ via the 1-parameter group of transformations $h_{\epsilon}$ from $\mathcal{M}_0$ to $\mathcal{M}_{\epsilon}$:
\begin{align}
Q_{\epsilon} &= h_{\epsilon \ast} \, \left[ Q_0 + \epsilon \left( \mathcal{L}_V Q_{\epsilon} \right)_{\epsilon = 0} + \mathcal{O} (\epsilon^2) \right] .
\end{align}
\n
The immediate implication here is that the first order term, the linearised perturbation of $Q$, will be gauge-invariant iff
\begin{align}
\mathcal{L}_V Q_0 = 0 ,
\end{align}
\n
for all 4-vectors $V$ on $\mathcal{M}_0$. This will be the case if:
\begin{itemize}
\item $Q_0$ vanishes in the background.
\item $Q_0$ is a constant scalar field.
\item $Q_0$ is some linear combination of Kronecker deltas with constant coefficients. 
\end{itemize}
\n
This is known as the Stewart-Walker lemma \cite{Stewart74} and provides a very powerful foundation from which gauge-invariant variables can be defined in the covariant approach to perturbation theory. This is due to the fact that in the covariant approach we aim to avoid any reference to a background spacetime and only invoke the background when determining which variables are zeroth order. Gauge-invariance of our set of equations is then guaranteed by the Stewart-Walker lemma if we find a complete set of variables that vanishes in the background. For the Schwarzschild spacetime, the only non-zero variables in the background were $\lbrace \calE, \phi , \calA \rbrace$ with all other scalars, vectors and tensors vanishing in the background due to symmetry. To construct gauge-invariant variables from these non-vanishing scalars, we just need to invoke spherical symmetry, which necessitates that the angular derivatives of the background scalars must vanish and will therefore constitute a set of gauge invariant variables. The set of variables which we will therefore work with in the linearised perturbation equations are as follows:
\begin{align}
\delta_a \lbrace \calE , \phi , \calA \rbrace &= \lbrace X_a , Y_a , Z_a \rbrace .
\end{align}
\n
In the formalism we have outlined above, we also have another implicit type of 'gauge' freedom, namely the freedom to choose a frame basis in the tangent space at each point. This is the same as in any other tetrad based approach, such as the Newman-Penrose formalism. As in \cite{Clarkson03}, we will use the term gauge-invariance to refer to invariance under a mapping between the true and background spacetimes and frame invariance to refer to invariance under our choice of frame vectors.

\subsection{Review: The Regge Wheeler Tensor}

In the metric approach to GR, a master equation for axial gravitational perturbations was originally derived by Regge and Wheeler \cite{Regge57} with the corresponding polar equation being derived by Zerilli \cite{Zerilli70} over a decade later. The structure of these wave equations can be shown to reduce to a Schr\"odinger like equation defined in terms of an effective potential $V_{\textrm{eff}}$. These original studies have formed the basis for many different approaches to studying linear perturbations to the Schwarzschild black hole \cite{Vishveshwara70,Price72,Moncrief74,Cunningham78,MartinGarcia00,Clarkson03,Martel05,Nagar05}. In the 1+1+2 approach to gravitational perturbations of a Schwarzschild black hole, the master variables for the polar and axial sectors can be unified into a single covariant, gauge invariant and frame invariant tensor $W_{\lbrace a b \rbrace}$ \cite{Clarkson03}. 

Building on this work, a master variable $W_{\lbrace a b \rbrace}$ for gravitational perturbations for the $f(R)$ Schwarzschild black hole has recently been derived in \cite{Nzioki14} and is given in terms of the shear 2-tensor $\zeta_{ab}$, a TT tensor constructed from the radial part of the electric Weyl tensor $\delta_{\lbrace a} X_{b \rbrace}$ and a TT tensor constructed from the curvature scalar $\delta_{\lbrace a} \delta_{b \rbrace} R$. The dimensionless, gauge-invariant, frame-invariant and transverse-traceless tensor $W_{ab}$ is defined by \cite{Nzioki14} \footnote{Note that \cite{Nzioki14} denotes this tensor as $M_{ab}$ whereas we use the notation of \cite{Clarkson03}.}
\begin{align}
 W_{ab} &= \frac{1}{2} \phi r^2 \zeta_{ab} - \frac{1}{3} \frac{r^2}{\calE} \delta_{\lbrace a} X_{b \rbrace} + \frac{f^{\prime \prime}_0}{3 f^{\prime}_0} r^2 \delta_{\lbrace a} \delta_{b \rbrace} R ,
 \label{eqn:RWTensor}
\end{align}
\n
where the even part of $W_{ab}$ is coupled to a curvature term: 
\begin{align}
\frac{1}{3} \frac{f^{\prime \prime}_0}{f^{\prime}_0} \delta_{\lbrace a} \delta_{b \rbrace} R = \frac{1}{3} \Pi_{ab}^R .
\end{align}
Consequentially, we have to include the trace equation for the curvature scalar in order for the wave equation to close. The curvature term vanishes in the axial sector and the variable $W_{ab}$ reduces to exactly the same form as in GR. The covariant wave equation this variable obeys is given by \cite{Nzioki14}
\begin{align}
 \ddot{W}_{\lbrace a b \rbrace} - \hat{\hat{W}}_{\lbrace a b \rbrace} - \calA \hat{W}_{\lbrace a b \rbrace} + \left( \phi^2 - \calE \right) W_{\lbrace a b \rbrace} - \delta^2 W_{\lbrace a b \rbrace} &= 0 ,
\end{align}
\n
which is identical to the covariant wave equation in GR \cite{Clarkson03,Pratten14}. Note that both the even and odd parity components of $W_{ab}$ obey the same wave equation. This tensor equation can be decomposed into scalar harmonics (see Appendix \ref{app:harmonics})
\begin{align}
 \ddot{W} - \hat{\hat{W}} - \calA \hat{W} + \left[ \frac{\ell (\ell + 1)}{r^2} + 3 \calE \right] W &= 0 ,
 \label{eqn:RWEqn}
\end{align}
\n
where $W = \lbrace W_T , \bar{W}_T \rbrace$. Associating the hat derivative with an affine parameter $\rho$, i.e. $\hat{\phantom{c}} = d / d \rho$, we can convert to the parameter $r$, $\rho \rightarrow r$, and then switch to the tortoise coordinates of Regge and Wheeler
\begin{align}
 r_{\ast} &= r + 2 M \ln \left( \frac{r}{2M} - 1 \right) .
\end{align}
\n
Letting 
\begin{align}
 \psi = \psi_{\textrm{RW}} = W ,
\end{align}
\n
we find that the harmonic equation reduces to a Schr\"odinger type equation
\begin{align}
 \left( \frac{d^2}{dr^2} + \sigma^2 \right) \psi &= V \psi ,
\end{align}
\n
with an effective potential $V_T$
\begin{align}
 V_T &= \left( 1 - \frac{2M}{r} \right) \, \left[ \frac{\ell ( \ell + 1 )}{r^2} - \frac{6 M}{r^3} \right] ,
\end{align}
\n
which is just the Regge-Wheeler (RW) potential for gravitational perturbations to the Schwarzschild spacetime \cite{Regge57,Clarkson03,Nzioki14}. 

\subsection{Review: Scalar Perturbations}
In FOG we note the presence of scalar modes that are not possible in Einstein GR. The trace equation yields a wave equation in terms of the Ricci scalar $R$. The equation obeys the same functional form as the generalised Regge-Wheeler equation for massive scalar perturbations in Einstein GR but with an effective mass given by \cite{Nzioki14}
\begin{align}
 U^2 &= \frac{f^{\prime}_0}{3 f^{\prime \prime}_0} .
\end{align}
\n
The covariant wave equation that the variable obeys is given by \cite{Nzioki14}
\begin{align}
 \ddot{\pazocal{R}} - \hat{\hat{\pazocal{R}}} - \calA \hat{\pazocal{R}} - \left( \calE - U^2 + \delta^2 \right) \pazocal{R} &= 0 , 
\end{align}
\n
where $\pazocal{R} = r \, R$. Introducing scalar spherical harmonics, this equation reduces to \cite{Nzioki14}
\begin{align}
 \ddot{\pazocal{R}}_S - \hat{\hat{\pazocal{R}}}_S - \calA \hat{\pazocal{R}}_S - \left[ \calE - \tilde{U}^2 - \frac{\ell (\ell + 1)}{r^2} \right] \pazocal{R}_S &= 0 ,
\end{align}
\n
where $\tilde{U} = C_1 / (3 C_2 )$, with $C_1$ and $C_2$ constants. Converting from the parameter $\rho$ to $r$ and then introducing tortoise coordinates, we find \cite{Nzioki14}
\begin{align}
 \left[ \frac{d^2}{d r_{\ast}^2} + \kappa^2 - V_S \right] \pazocal{R} &= 0 ,
\end{align}
\n
where
\begin{align}
 V_S &= \left( 1 - \frac{2 M}{r} \right) \, \left[ \frac{\ell (\ell + 1)}{r^2} + \frac{2M}{r^2} + \tilde{U}^2 \right] .
\end{align}

\section{The Weyl Terms}\label{sec:WTMain}
We now move on to the core subject of the paper, namely the description of gravitational perturbations via the Weyl tensor in the 1+1+2 formalism. The definition of a master variable for gravitational perturbations is non-unique (see e.g. \cite{Nagar05}) and different definitions will provide different advantages or insight. The previous definition of the RW tensor $W_{ab}$ given in \cref{eqn:RWTensor} allows us to interpret the RW equation as being related to the fluctuations in the radial part of the electric Weyl tensor as well as the distortions of the sheet, given by $\zeta_{ab}$. Alternatively, we may directly invoke the decomposition of the Weyl tensor to introduce a set of new master variables. This allows us to relate the 1+1+2 RW tensor to $h_{+}$ and $h_{\times}$, the NP Weyl scalars and the 2+2 gauge invariant master variables in a clear, transparent manner. 

\subsubsection{Unified Polar and Axial Gravitational Perturbations}
It is a well established fact that gravitational waves propagate in a perturbed Schwarzschild spacetime and it would be reasonable to expect that the transverse-traceless tensors governing the gravitational perturbations will obey wave equations that reduce to the plane wave case in the limit $M \rightarrow 0$. The most natural variables to study would be those relating to the transverse-traceless (TT) degrees of freedom in the Weyl curvature tensor, as this provides a description of the free gravitational field. In the 1+1+2 formalism, the TT components of the electric and magnetic Weyl tensor will just be the 2-tensors $\mathcal{E}_{ab}$ and $\mathcal{H}_{ab}$. Constructing the wave equations for these variables, however, we find that they do not form a closed system, they contain forcing terms from other 1+1+2 tensors. Such couplings are not present in the plane wave limit making the interpretation and solution of these wave equations in isolation non-trivial. 

In order to find a wave equation for the Weyl curvature 2-tensors, we need to perform a 1+1+2 decomposition of the 1+3 evolution equations for the Weyl curvature (see Eqns. 1.3.42 and 1.3.43 in \cite{Tsagas07}). Linearising about the Schwarzschild background, this yields \cite{Clarkson03,Nzioki14}: 
\begin{align}
\dot{\calE}_{\lbrace a b \rbrace} - \epsilon_{c \lbrace a} \hat{\calH}_{b \rbrace}^{\phantom{b \rbrace} c} &= - \epsilon_{c \lbrace a} \delta^c \calH_{b \rbrace} + \left( \frac{1}{2} \phi + 2 \calA \right) \epsilon_{c \lbrace a} \calH_{b \rbrace}^{\phantom{b \rbrace} c} - \frac{3}{2} \calE \Sigma_{ab} , \label{eqn:HWeylPropEv}  \\
\dot{\calH}_{\lbrace a b \rbrace} + \epsilon_{c \lbrace a} \hat{\calE}_{b \rbrace}^{\phantom{b \rbrace} c} &= \epsilon_{c \lbrace a} \delta^c \calE_{b \rbrace} + \frac{3}{2} \calE \epsilon_{c \lbrace a} \zeta_{b \rbrace} - \left( \frac{1}{2} \phi + 2 \calA \right) \epsilon_{c \lbrace a} \calE_{b \rbrace}^{\phantom{b \rbrace} c} .
\label{eqn:EWeylPropEv}
\end{align}
\n
The covariant wave equations for the electric and magnetic Weyl 2-tensors can now be derived by applying the wave operator $\ddot{\Psi} - \hat{\hat{\Psi}}$ to $\calE_{ab}$ and $\calH_{ab}$ using the above expressions to yield
\begin{align}
  &\ddot{\calH}_{\lbrace a b \rbrace} - \hat{\hat{\calH}}_{\lbrace a b \rbrace} - \left( \phi + 5 \calA \right) \hat{\calH}_{\lbrace a b \rbrace} - \delta^2 \calH_{\lbrace a b \rbrace}  + \left[ \frac{1}{2} \phi^2 - 5 \calE \right] \calH_{\lbrace a b \rbrace} = \left[ - 3 \calE \, \epsilon_{\lbrace a} \Sigma_{b \rbrace}^{\phantom{\rbrace b} c} + 2 \delta_{\lbrace a} \calH_{b \rbrace} \right] \left( \phi - 2 \calA \right) \label{eqn-HWave} \\
 &\ddot{\calE}_{\lbrace a b \rbrace} - \hat{\hat{\calE}}_{\lbrace a b \rbrace} - \left( \phi + 5 \calA \right) \hat{\calE}_{\lbrace a b \rbrace} - \delta^2 \calE_{\lbrace a b \rbrace} + \left[ \frac{1}{2} \phi^2 - 5 \calE \right] \calE_{\lbrace a b \rbrace} = \left[ 3 \calE  \zeta_{\lbrace a b \rbrace} + 2 \delta_{\lbrace a} \calE_{b \rbrace} \right] \left( \phi - 2 \calA \right) .  \label{eqn-EWave}
\end{align}
\n
Clearly, these wave equations are coupled to the 1+1+2 tensors $\Sigma_{\lbrace a b \rbrace}$, $\zeta_{\lbrace a b \rbrace}$, $\mathcal{E}_a$ and $\mathcal{H}_a$. Neither of these variables alone will obey a closed system of equations. By inspection of the propagation and evolution equation we can start to see similarities in the structures of these two wave equations. Importantly, the coupling terms can be on the RHS of Eqns. (\ref{eqn:HWeylPropEv}) and (\ref{eqn:EWeylPropEv}) share similar pre-factors that are reflected in the RHS of Eqns. (\ref{eqn-HWave}) and (\ref{eqn-EWave}) suggesting that we should be able to decouple these 2-tensors from the other 1+1+2 variables. 

Following this insight, we can construct a new variable from the linear combination of the electric and magnetic Weyl 2-tensors $\pazocal{J}^{\pm}_{ab}$ that completely decouples from the other 1+1+2 variables. Explicitly, this new perturbation variable is given by
\begin{align}
 \pazocal{J}^{\pm}_{\lbrace a b \rbrace} &= \calE_{\lbrace a b \rbrace} \pm \epsilon_{c \lbrace a} \calH_{b \rbrace}^{ \phantom{ b \rbrace } c} , \\
 &= \left[ \calE_T \mp \bar{\calH}_T \right] Q_{ab} + \left[ \bar{\calE}_T \pm \calH_T \right] \bar{Q}_{ab} ,
 \label{eqn:WeylMaster}
\end{align}
\n
where in the second line we have decomposed the perturbation variable into tensorial harmonics as per Appendix \ref{app:harmonics}. More importantly, this variable can be shown to obey the following closed, covariant and gauge invariant wave equation
\begin{align}
 \ddot{\pazocal{J}}^{\pm}_{\lbrace a b \rbrace} - \hat{\hat{\pazocal{J}}}^{\pm}_{\lbrace a b \rbrace} - \left( \calA + 3 \phi \right) \hat{\pazocal{J}}^{\pm}_{\lbrace a b \rbrace} \mp \left( 4 \calA - 2 \phi \right) \dot{\pazocal{J}}^{\pm}_{\lbrace a b \rbrace} - \left( \delta^2 + 2 K \right) \pazocal{J}^{\pm}_{\lbrace a b \rbrace} + \left( 4 \calA^2 - 4 \calE \right) \pazocal{J}^{\pm}_{\lbrace a b \rbrace} &= 0 ,
 \label{eqn:WeylMasterWave}
\end{align}
\n
which, when decomposed into covariant harmonics, reduces to
\begin{align}
 \ddot{\pazocal{J}}^{\pm} - \hat{\hat{\pazocal{J}}}^{\pm} - \left( \calA + 3 \phi \right) \hat{\pazocal{J}}^{\pm} \mp \left( 4 \calA - 2 \phi \right) \dot{\pazocal{J}}^{\pm} + \left[ \frac{\ell (\ell + 1)}{r^2} - \frac{3}{2} \phi^2 + 2 \calE + 4 \calA^2 \right] \pazocal{J}^{\pm} &= 0 ,
\end{align}
\n
where $\pazocal{J}^{\pm} = \lbrace \pazocal{J}^{\pm}_T , \bar{\pazocal{J}}^{\pm}_T \rbrace$. As with $W_{\lbrace ab \rbrace}$, this variable unifies the even and odd parity perturbations into a single transverse-traceless tensor $\pazocal{J}_{ab}$. By construction, this variable is defined by the transverse-traceless degrees of freedom in the electric and magnetic Weyl tensors. It is therefore natural that this variable should describe the propagation of gravitational waves in the Schwarzschild spacetime. In the limit $M \rightarrow 0$, we recover the plane wave case, as expected. Additionally, we note that gauge invariance of our perturbation variable is guaranteed by the Stewart-Walker lemma \cite{Stewart74} as all vectors and tensors vanish in the background spacetime. 

In the context of $f(R)$ gravity, it may seem a little curious that the wave equation for this perturbation variable is exactly the same as found in General Relativity. However, this result should not come as too much of a surprise. The perturbation variable, being TT in nature, explicitly describes massless modes that propagate along null curves. The scalar modes, having an effective mass of $U^2$, will physically propagate along timelike curves on the black hole background. In addition, we are explicitly considering a constrained class of solutions in $f(R)$ for which the scalar modes are not excited in the background spacetime. As such, the scalar modes and tensor modes decouple at linear order. This decoupling between the massive and massless modes has previously been noted \cite{Myung11,Nzioki14} but the derivation via metric based approaches can often obfuscate the underlying physics. By adopting the geometrically and physically meaningful 1+1+2 semi-tetrad approach, we can explicitly see which kinematical and gravitational terms are of importance and we can isolate physically meaningful variables, such as the TT components of the electric and magnetic Weyl tensors. 

As a side note, by geometrically and physically meaningful we are referring to the notion that everything can be related to the dynamics and kinematics of congruences as well as the fact that all objects are gauge-invariant and covariantly defined without reference to a given coordinate system. The derivative operators appearing in the wave equation are just convective derivatives along the timelike and spacelike congruences or angular derivatives on the 2-surface. The perturbation variable itself tells us about the evolution of the Weyl 2-tensors that are defined on the 2-surfaces and therefore provides a natural description of gravitational radiation. In addition, these variables can be transparently related to the broader class of LRS-II spacetimes, where it can be shown that the perturbation variable defined above describes gravitational perturbations to vacuum LRS-II spacetimes \cite{PrattenThesis}. 

Finally, it is also worth noting that were we to consider more astrophysical realistic systems or systems with matter content, the role of the scalar modes can become pronounced. An example of such behaviour can be seen in the dynamical secularisation of isolated neutron stars in the late-inspiral and merger phases of a compact binary coalescence \cite{Barausse13}. A study of neutron star systems in the 1+1+2 formalism would provide a natural extension to this work. 

\subsubsection{Axial Gravitational Perturbations}\label{sec:112Axial}
As mentioned previously, the curvature term only occurs in the even parity sector. This means that axial gravitational perturbations to the $f(R)$ Schwarzschild black hole will be governed by the same covariant wave equation as in General Relativity. A natural candidate for such a perturbation variable is the magnetic Weyl tensor. Consequentially, we introduce a perturbation variable $\pazocal{V}_{\lbrace a b \rbrace}$ constructed from the radial part of the magnetic Weyl tensor
\begin{align}
 \pazocal{V}_{\lbrace a b \rbrace} &= r^2 \, \delta_{\lbrace a} \delta_{b \rbrace } \calH .
 \label{eqn:WeylMasterAxial}
\end{align}
\n
This variable can be shown to obey the following covariant wave equation
\begin{align}
 \ddot{\pazocal{V}}_{\lbrace a b \rbrace} - \hat{\hat{\pazocal{V}}}_{\lbrace a b \rbrace} - \left( \calA + 3 \phi \right) \hat{\pazocal{V}}_{\lbrace a b \rbrace} - \left[  \delta^2 + 2 K \right] \pazocal{V}_{\lbrace a b \rbrace} &= 0 .
 \label{eqn:WeylMasterWaveAxial}
\end{align}
\n
Decomposing the above equation into tensor harmonics\footnote{As per Appendix \ref{app:harmonics}} we find that
\begin{align}
 \ddot{\pazocal{V}} - \hat{\hat{\pazocal{V}}} - \left( \calA + 3 \phi \right) \hat{\pazocal{V}} + \left[ \frac{\ell \left( \ell + 1 \right)}{r^2} - \frac{3}{2} \phi^2 + 6 \calE \right] \pazocal{V} &= 0 .
 \label{eqn:WeylMasterWaveAxialHarmonic}
\end{align}
Though this harmonic equation looks different to the Regge-Wheeler equation, this variable is related to Regge-Wheeler variable. This can be seen by rescaling the tensor defined above as follows: $\pazocal{X} = r^3 \pazocal{V}$,  \cref{eqn:WeylMasterWaveAxialHarmonic} then reduces to
\begin{align}
 0 &= \frac{1}{r^3} \left[ \ddot{\pazocal{X}} - \hat{\hat{\pazocal{X}}} - \calA \hat{\pazocal{X}} + \left\lbrace \frac{\ell (\ell + 1)}{r^2} + 3 \calE \right\rbrace \pazocal{X} \right] \\
 &= \ddot{\pazocal{X}} - \hat{\hat{\pazocal{X}}} - \calA \pazocal{X} + \left\lbrace \frac{\ell (\ell + 1)}{r^2} + 3 \calE \right\rbrace \pazocal{X} .
\end{align}
\n
This is nothing more than the Regge-Wheeler equation as per \cref{eqn:RWEqn}. This allows us to reinterpret the RW master variable as simply the radial part of the magnetic Weyl scalar as $\calH = n^a n^b H_{ab}$. This is more physically intuitive than the original derivation via metric perturbation theory and is manifestly covariant and gauge invariant. 

\section{Relation to Other Formalisms}\label{sec:OtherFormalism}
\subsection{The Newman-Penrose Formalism}
Another way to interpret our new perturbation variables is to re-express the Newman-Penrose scalars in terms of the 1+1+2 variables. The Newman-Penrose formalism is a full tetrad approach in which the underlying frame vectors form a null tetrad consisting of two real null vectors $( l^a , k^a)$ and a complex-conjugate pair $(m^a , \bar{m}^a)$ \cite{Newman62}. The two real null vectors correspond to ingoing and outgoing null congruences, whereas the complex-conjugate pair will simply correspond to a decomposition of the 2-surface orthogonal to $u^a$ and $n^a$. The frame vectors are defined as follows:
\begin{align}
l_a &= \frac{1}{\sqrt{2}} \left( u_a + n_a \right) \qquad && l_a l^a = 0\\
k_a &= \frac{1}{\sqrt{2}} \left( u_a - n_a \right) \qquad && k_a k^a = 0 \qquad l_a k^a = -1\\
m_a &= \frac{1}{\sqrt{2}} \left( v_a - i w_a \right) \qquad && m^a \bar{m}_a = 1 \qquad m^a m_a = \bar{m}^a \bar{m}_a = 0, 
\end{align}
\n
with the metric being decomposed as follows
\begin{align}
g_{ab} &= - l_a k_b - k_a l_b + 2 m_{(a} \bar{m}_{b)} \\
&= -u_a u_b + n_a n_b + N_{ab},
\end{align}
\n
where we have identified the induced 2-metric with the two complex frame vectors: $N_{ab} = 2 m_{(a} \bar{m}_{b)}$. The Newman-Penrose scalars are defined by the appropriate contractions of the Weyl tensor with respect to this null tetrad and are given by\footnote{Note that other definitions or conventions exist.}
\begin{align}
\Psi_0 &= C_{abcd} l^a m^b l^c m^d \\
\Psi_1 &= C_{abcd} l^a m^b l^c k^d \\
\Psi_2 &= C_{abcd} l^a m^b \bar{m}^c k^d \\
\Psi_3 &= C_{abcd} l^a k^b \bar{m}^c k^d \\
\Psi_4 &= C_{abcd} \bar{m}^a k^b \bar{m}^c k^d ,
\end{align}
\n
with the 5 complex Weyl scalars encoding the 10 independent components of the Weyl tensor. The conventional physical intuition applied to these scalars is as follows: $\Psi_0$ describes transverse radiation along $k^a$ and thereby describes ingoing gravitational radiation. $\Psi_1$ is an ingoing gauge wave described by the longitudinal radiation along $k^a$. $\Psi_2$ is a Coulomb term related to gravitation attraction and frame dragging effects. $\Psi_3$ describes longitudinal radiation along $l^a$ and is therefore an outgoing gauge wave. Finally, $\Psi_4$ describes transverse radiation along $l^a$ and is thereby characterises outgoing gravitational radiation. 

One of the motivations for considering the Weyl scalars is that the response of a GW detector will be encoded in these curvature scalars in the Jordan frame \cite{Eardley74}. In particular, for outgoing waves, the Weyl peeling theorem tells us how the Weyl scalars fall off along outgoing radial null geodesics in some neighbourhood of future null infinity $\cal{J}^{+}$:
\begin{align}
\Psi_n \sim r^{n-5} .
\end{align}
\n
This result is relatively robust and valid for a wide range of tetrad frames given a suitable choice for $r$. In an asymptotically flat spacetime, given that we are sufficiently far away from the source, the outgoing waves can be expressed within the realms of linearised gravity in the TT gauge:
\begin{align}
\Psi_0 = \Psi_1 = \Psi_2 = \Psi_3 = 0 \qquad \Psi_4 = - \ddot{h}_{+} + i \ddot{h}_{\times} ,
\end{align}
\n
with $h_{+}$ and $h_{\times}$ the standard gravitational wave polarisations, i.e the 2 graviton degrees of freedom. This means that in General Relativity, the radiative degrees of freedom that decay as $r^{-1}$ and are observable by GW detectors are simply encoded in $\Psi_4$. The extension to $f(R)$ gives us the possibility of exciting an additional transverse, radiative scalar mode. However, under many general considerations, the radiative component of this scalar mode tends to vanish due to constraints on the underlying scalar-tensor theory \cite{Barausse13}. As such, the scalar modes will only couple weakly to a GW detector making direct detections difficult \cite{Damour98,Barausse13}.  

Performing a systematic decomposition of all the Weyl tensor into it's constituent 1+1+2 components, the Weyl scalars can be re-written as follows \cite{StephaniBook}:
\begin{align}
\Psi_0 &= \left[ \mathcal{E}_{ab} + \epsilon_{r \lbrace a} \mathcal{H}_{b \rbrace}^{\phantom{b \rbrace} r} \right]  \, m^a m^b  \\
\Psi_1 &= - \frac{1}{\sqrt{2}} \, \left[ \mathcal{E}_a - \epsilon_{ab} \mathcal{H}^b \right] m^a \\
\Psi_2 &= \frac{1}{2} \left[ \mathcal{E} - i \mathcal{H} \right] \\
\Psi_3 &= \frac{1}{\sqrt{2}} \, \left[ \mathcal{E}_a + \epsilon_{a b} \mathcal{H}^b \right] \, \bar{m}^a \\
\Psi_4 &= \left[ \mathcal{E}_{ab} - \epsilon_{r \lbrace a} \mathcal{H}_{b \rbrace}^{\phantom{b \rbrace} r} \right] \, \bar{m}^a \bar{m}^b  .
\end{align}
\n
This decomposition follows from analogous results derived in a 3+1 or 1+3 approach \cite{Barnes89,StephaniBook,AlcubierreBook}.
As is usual in a tetrad formalism, these scalars will, in general, be frame-dependent with the behaviour of these scalars under transformations explicitly known \cite{StewartBook}. 
Reassuringly, the only Newman-Penrose scalar that is non-vanishing in the background spacetime is $\Psi_2$ for which we recover:
\begin{align}
\Psi_2 &= \frac{1}{2} \calE = - \frac{M}{r^3} ,
\end{align}
\n
as per the parametric solution for the 1+1+2 variables in \cref{eqn:bgpara}. The above results are also consistent with the literature in the sense that all LRS spacetimes will be of Petrov type D or O \cite{Ellis99}\footnote{See Appendix \ref{app:Petrov} for a brief review of the Petrov classification.}. This can be seen by the simple requirement that under LRS, all 2-vectors and 2-tensors must vanish in the background. This leaves $\Psi_2 \neq 0$ with the Petrov type determined by $\calE$ and $\calH$. If these scalars vanish the Petrov type is O, otherwise the Petrov type is D. Under perturbations, the Petrov type will be asymptotically type N due to the Weyl peeling theorem. We can extend this discussion to include the four non-zero vacuum invariants constructed from the Riemann curvature tensor \cite{AlcubierreBook}
\begin{align}
I &= 3 \Psi^2_2 - 4 \Psi_1 \Psi_3 + \Psi_0 \Psi_4 , \\
J &= \Psi_0 \Psi_2 \Psi_4 + 2 \Psi_1 \Psi_2 \Psi_3 - \Psi_0 \Psi^2_3 - \Psi^2_1 \Psi_4 - \Psi^3_2 . 
\end{align}
\n
In a Petrov type D spacetime, as Petrov type O is trivial, these invariants reduce to 
\begin{align}
I = 3 \Psi^2_2 \qquad \textrm{and} \qquad J = - \Psi^3_2 . 
\end{align}
\n
From our decomposition of the NP scalars, these can be expressed via the 1+1+2 scalars as
\begin{align}
I &= \frac{3}{4} \left[ \calE^2 - \calH^2 - 2 i \calE \calH \right] , \\
J &= - \frac{1}{8} \left[ \calE^2 - 3 i \calE^2 \calH - 3 \calE \calH^2 + i \calH^3 \right] , \\
S &= 27 \frac{J^2}{I^3} = 1 \qquad \textrm{For Type D or Type II Spacetimes.}
\end{align}

From these decompositions we can immediately identify the correspondence between the perturbation variable $\pazocal{V}_{ab}$ and the imaginary part of $\Psi_2$:
\begin{align}
\Im{\left[r^3 \Psi_2\right]} \sim \psi_{RW} \sim r^3 \calH .
\end{align}
\n
The relationship between the imaginary part of $\Psi_2$ and the Regge-Wheeler variable was first discussed by \cite{Price72} \footnote{Note that \cite{Price72} uses a different notation for the NP scalars based on their spin-weight: $\lbrace \Psi_{-2} , \Psi_{-1}, \Psi_0, \Psi_1 , \Psi_2 \rbrace$.}. 

Similarly, we can immediately identify a correspondence between $\lbrace \Psi_0 , \Psi_4 \rbrace$ and the perturbation variable $\pazocal{J}^{\pm}_{ab}$. Notably, we see that 
\begin{align}
\Psi_0 = \pazocal{J}^{+}_{ab} \; m^a m^b , \\
\Psi_4 = \pazocal{J}^{-}_{ab} \; \bar{m}^a \bar{m}^b .
\end{align}
\n
This allows us to associate $\pazocal{J}^{-}_{ab}$ with outgoing gravitational radiation\footnote{See Appendix \ref{app:Energy}.} and reinforces the notion that $\pazocal{J}^{\pm}_{ab}$ contains, in a compact form, the gravitational wave polarisations $\lbrace h_{+} , h_{\times} \rbrace$. 

By construction, the scalar degrees of freedom present in the class of $f(R)$ theories considered here do not enter the Weyl scalars. This is reflected in our ability to both construct a closed, covariant wave equation that decouples from the scalar modes as well as the inability of the scalar modes to affect the potential of the tensor degrees of freedom at linear order. 

A useful approach for studying the curvature-fluid perturbations is via the Ricci scalars. In the most general case, these scalar quantities will encode the ten independent components of the Ricci tensor in the form of three real scalars $\lbrace \Phi_{00} , \Phi_{11} , \Phi_{22} \rbrace$ and six complex scalars $\lbrace \Phi_{01} = \bar{\Phi}_{10} , \Phi_{02} = \bar{\Phi}_{20} , \Phi_{12} = \bar{\Phi}_{21} \rbrace$. Unlike the Weyl scalars, the Ricci scalars are implicitly related to the energy-momentum distribution of the spacetime via EFE. In the context of perturbations to the $f(R)$ Schwarzschild black hole, these scalars will measure excitations of the curvature fluid. The Ricci scalars are defined as follows \cite{StephaniBook}:
\begin{align}
\Phi_{00} &= \frac{1}{2} R_{ab} l^a l^b , \\
\Phi_{11} &= \frac{1}{4} R_{ab} \left( l^a k^b + m^a \bar{m}^b \right) , \\
\Phi_{22} &= \frac{1}{2} R_{ab} k^a k^b , \\
\Phi_{01} &= \bar{\Phi}_{10} = - \frac{1}{2} R_{ab} l^a m^b , \\
\Phi_{02} &= \bar{\Phi}_{20} = \frac{1}{2} R_{ab} m^a m^b  \\
\Phi_{12} &= \bar{\Phi}_{21} =  \frac{1}{2} R_{ab} \bar{m}^a k^b .
\end{align}
\n
As per the Weyl scalars, we can systematically decompose the Ricci scalars into their constituent 1+1+2 variables. We find the following:
\begin{align}
\Phi_{00} &= \frac{1}{2} \left( p - Q + \frac{1}{2} \Pi \right) , \\
\Phi_{11} &= \frac{1}{4} \left( \mu - \Pi + \Pi_{ab} m^a \bar{m}^b \right)  , \\
\Phi_{22} &= \frac{1}{2} \left( p + Q + \frac{1}{2} \Pi \right) , \\
\Phi_{01} &= \frac{1}{2 \sqrt{2}} \left( \Pi_a - Q_a \right) m^a , \\
\Phi_{02} &= \frac{1}{2} \Pi_{ab} m^a m^b , \\
\Phi_{12} &= - \frac{1}{2 \sqrt{2}} \left( \Pi_a + Q_a \right) \bar{m}^a .
\end{align}
\n
The 1+1+2 expressions presented above are valid in both General Relativity as well as the $f(R)$ extension considered in this paper. In the Schwarzschild background, these variables naturally vanish as we only consider vacuum gravitational perturbations, i.e. $T^M_{ab}$ vanishes to all orders. In the $f(R)$ extension however, these variables will vanish in the background but will be non-zero once we consider first-order perturbations involving the scalar degrees of freedom. For example, we can explicitly insert the linearised energy-momentum tensor for the curvature fluid into $\Phi_{00}$  and $\Phi_{22}$ to obtain:
\begin{align}
\Phi_{00} &= \frac{f^{\prime \prime}_0}{2 f^{\prime}_0} \left[ \ddot{R} - \frac{1}{3} \hat{\hat{R}} - \hat{R} \left( \mathcal{A} + \frac{5}{6} \phi \right) - \frac{5}{6}  \delta^2 R +  \dot{\hat{R}} -  \mathcal{A} \dot{R}  \right] , \\
\Phi_{22} &= \frac{f^{\prime \prime}_0}{2 f^{\prime}_0} \left[ \ddot{R} + \frac{1}{3} \hat{\hat{R}} - \hat{R} \left( \calA + \frac{5}{6} \phi \right) - \frac{5}{6} \delta^2 R - \frac{5}{3} \hat{\dot{R}} \right] .
\end{align}

\subsection{The 2+2 Formalism}
The other approach to which we wish to make a connection is that of the 2+2 formalism. This formalism was originally introduced in \cite{Gerlach79,Gerlach80} and a completion of this formalism was given by \cite{Gundlach00} who providing a systematic derivation of the gauge-invariants, introduced a fluid-frame decomposition and wrote down the resulting system of master equations governing gravitational perturbations of spherically symmetric spacetimes. The approach is built around the decomposition of the background 4-dimensional spacetime $\mathcal{M}^4$ into a  warped product $\mathcal{M}^4 = \mathcal{M}^2 \otimes S^2$, where $\mathcal{M}^2$ is a 2-dimensional Lorentzian manifold and $S^2$ is the 2-sphere. In essence, we are covariantly factorising out the spherical symmetry and reducing the problem to a two-dimensional problem written in terms of the two essential coordinates: time and radius. As such, the metric can be written as\footnote{Note carefully: In this section, $\lbrace A, B, \dots \rbrace$ denote coordinates on $\mathcal{M}^2$ and $\lbrace a, b, \dots \rbrace$ denote coordinates on $S^2$.}
\begin{align}
ds^2 = g_{AB} (x^C) dx^a dx^B + r^2 (x^C) \gamma_{ab} dx^a dx^b ,
\end{align}
\n
where $g_{AB}$ is the metric on $\mathcal{M}^2$ and $\gamma_{ab}$ is the metric on the unit sphere $S^2$. The scalar $r = r(X^a)$ is defined on $\mathcal{M}^2$ and can be identified as the invariantly defined radial coordinate of spherically-symmetric spacetimes. Using this decomposition, the EFE in vacua can be re-expressed as
\begin{align}
G_{AB} &= - 2 \left( v_{A | B} + v_A v_B \right) + 2 \left( v_C^{\phantom{C} |C} + 3 v_C v^C - \frac{1}{r^2} \right) g_{AB} = t_{AB} , \\
\frac{1}{2} G^a_{\phantom{a} a} &= v_C^{\phantom{C} |C} + v_C v^C - \mathcal{R} = Q, \\
\nonumber v_A &= \frac{r_{|A}}{r} ,
\end{align}
\n
where $\mathcal{R} = \frac{1}{2} R^A_{\phantom{A}}$ is the Gaussian curvature of $\mathcal{M}^2$. The energy-momentum conservation equations take the following form:
\begin{align}
t_{AB}^{\phantom{AB} | B} + 2 t_{AB} v^B - 2 v^A Q &= 0 .
\end{align}

Following \cite{Gundlach00}, all metric perturbations can be written as a scalar, vector or tensor field on $\mathcal{M}^2$ times a spherical harmonic scalar, vector or tensor field on $S^2$. As such, the axial metric perturbations can be written as:
\begin{align}
 h^{\textrm{Axial}}_{\mu \nu} &= \left( 
 	  \begin{array}{ll}
          0 & h^{\textrm{Axial}}_A \, \bar{Y}_a \\
         h^{\textrm{Axial}}_A \, \bar{Y}_a & h \bar{Y}_{a b}
         \end{array} \right) ,
\end{align}
\n
and the polar metric perturbations as:
\begin{align}
 h^{\textrm{Polar}}_{\mu \nu} &= \left( 
 	  \begin{array}{ll}
          h_{AB} \, Y & h^{\textrm{Polar}}_A \, {Y}_a \\
         h^{\textrm{Polar}}_A \, {Y}_a & r^2 \left( K \, Y \, \gamma_{a b} + G {Y}_{: a b} \right)
         \end{array} \right) .    
\end{align}
\n
Note that there is a natural correspondence between the covariant harmonics $\lbrace Q , Q_a , Q_{ab} \rbrace$ and their spherical harmonic counterparts $\lbrace Y , Y_a , Y_{ab} \rbrace$, this is discussed in Appendix \ref{app:harmonics}. In practice it is often convenient to adopt the Regge-Wheeler gauge $\lbrace h, h^{\textrm{Polar}}_A , G \rbrace = 0$ as, in this gauge, the remaining gauge-invariant variables simply correspond one-to-one with the bare perturbations. Knowing how the variables transform under general gauge-transformations, we could always map to a different gauge using known algebraic relations. 

\subsubsection{Polar Perturbations and Correspondence}
Using a radial-frame in $\mathcal{M}^2$, the metric perturbation $k_{AB}$ can be split into three gauge-invariant scalars $\lbrace \chi , \varphi,  \varsigma \rbrace$
\begin{align}
k_{AB} &= \left( \chi + \varphi \right) \left( n_A n_B + u_A u_B \right) + \varsigma \left( u_A n_B + n_A u_B \right) .
\label{eq.FluidFramekAB}
\end{align}
\n
Substituting this decomposition into the 2+2 EFE, we can recover a system of scalar equations by the appropriate projection operations with respect to the basis $\lbrace u, n \rbrace$. The system of master equations for the Schwarzschild spacetime in GR can be written as follows \cite{Gundlach00,Nagar05}\footnote{Note that in this section a dot derivative is defined by $u^A \tilde{\nabla}_A$ and a prime derivative by $n^A \tilde{\nabla}_A$, where $\tilde{\nabla}_A$ is the covariant derivative on $\mathcal{M}^2$, i.e. $\tilde{\nabla}_C g_{AB} = 0$.}
\begin{align}
-\ddot{\chi} + \chi^{\prime \prime} &= -2 \left[ 2 \nu^2 - \frac{6 M}{r^3} \right] \left( \chi + \varphi \right) - \left( 5 \nu - 2 W \right) \chi^{\prime} + \frac{(\ell + 2)(\ell - 1)}{r^2} \chi , \label{eqn:2p2Master} \\
-\ddot{\varphi} &= - W \chi^{\prime} - \nu \varphi^{\prime} - \frac{4 M}{r^3} \left( \chi + \varphi \right) - \frac{(\ell + 2)(\ell - 1)}{2 r^2} \chi , \nonumber \\
-\dot{\varsigma} &= 2 \nu \left( \chi + \varphi \right) + \chi^{\prime} . \nonumber
\end{align}
\n
In addition, we also have three constraint equations on the Cauchy data for $\lbrace \chi, \varphi , \varsigma \rbrace$:
\begin{align}
\left( \dot{\varphi} \right)^{\prime} &= - W \dot{\chi} + \frac{\ell (\ell + 1)}{2 r^2} \varsigma , \\
\varphi^{\prime \prime} &= \frac{\ell (\ell + 1)}{r^2} \left( \chi + \varphi \right) - \frac{(\ell + 2)(\ell - 1)}{2 r^2} \chi + W \chi^{\prime} - 2 W \varphi^{\prime} , \\
\varsigma^{\prime} &= - 2 \nu \varsigma - \dot{\chi} - 2 \dot{\varphi} .
\end{align}
\n
Clearly, from the above, the highest derivatives of $\chi$ form a wave equation with characteristics set by the metric $g_{\mu \nu}$ and Cauchy data $\lbrace \chi , \dot{\chi} \rbrace$ that can be set independently of matter perturbations \cite{Gundlach00}. It was therefore noted in \cite{Gundlach00} that $\chi$ can be reasonably said to characterise polar gravitational waves. The interpretation of $\varphi$ is a less trivial but was originally noted to correspond to longitudinal gravitational waves made physical by the presence of matter. As we work in a vacuum spacetime, how we should interpret $\varphi$ is not clear. Finally, $\varsigma$ was noted to be a term that is advected with the fluid. These physical descriptions of the gauge-invariants can be made manifest by explicitly writing down a correspondence between the 1+1+2 variables and the 2+2 gauge invariant perturbations. In essence, this approach to understanding the physical and geometrical meaning of gauge invariant perturbations is reminiscent of \cite{Bruni92} but adapted to spherically symmetric spacetimes. 

Performing a systematic study of the 1+1+2 variables in terms of the 2+2 gauge invariants, we find the that the gauge-invariant perturbations are captured by the electric and magnetic Weyl 2-tensors:
\begin{align}
\mathcal{E}_{ab} &= - \frac{1}{2} \left( \chi + \varphi \right) \, Y_{ab} , \\
\mathcal{H}_{ab} &=  - \frac{1}{2} \varsigma \, Y_{ab} .
\end{align}
\n
In many ways this is highly reassuring. After all, it was noted that $\chi$ and $\varphi$ somehow relate to polar gravitational wave degrees of freedom. Similarly, the interpretation of $\varsigma$ was less transparent but, via the 1+1+2 approach, we can clearly see that this gauge-invariant perturbation is related to the magnetic Weyl 2-tensor and therefore will be related to genuine relativistic effects, such as frame-dragging. This also corresponds to the the result presented earlier in \cref{eqn:WeylMasterWave} whereby the TT degrees of freedom of the Weyl tensor were shown to obey a closed covariant wave equation. As such, the master variable $\pazocal{J}_{\lbrace a b \rbrace}$ and its concomitant wave equation are simply a covariant, gauge-invariant and frame-invariant repackaging of the system of equations given by \cref{eqn:2p2Master} for the gauge-invariant perturbations $\lbrace \chi, \varphi , \varsigma \rbrace$. 

From the result in \cref{eqn:WeylMasterWave}, which is valid for both General Relativity as well as $f(R)$, it would be natural to assume that were a 2+2 decomposition of the $f(R)$ field equations to be performed, an analogous closed system of equations for these gauge invariant perturbations could be found. Implicitly, we would also have to fold in first-order energy momentum perturbations corresponding to excitations of the scalar degrees of freedom. This would follow a procedure analogous to that used in the 1+1+2 decomposition and linearisation of the energy-momentum tensor for the curvature fluid. 

As a final point, we can complete the correspondence between the 1+1+2, 2+2 and NP polar sector perturbations by studying the perturbations to $\Psi_4$ in terms of the 2+2 gauge-invariant variables. In the background spacetime, $\Psi_4$ vanishes and therefore its perturbation is naturally gauge-invariant. This perturbed Weyl scalar can be written as
\begin{align}
\delta \Psi_4 &= \delta C_{abcd} \bar{m}^a k^b \bar{m}^c k^d \\
&= \delta R_{a B c D} \bar{m}^a \bar{m}^c k^B k^D \\
&= - \frac{1}{2} k^B k^D \, k_{BD} \, \left[ \bar{m}^a \bar{m}^c Y_{ac} \right] \\
&= - \frac{1}{2} \left[ \chi + \varphi + \varsigma \right] \left[ \bar{m}^a \bar{m}^c Y_{ac} \right] \\
&\sim \pazocal{J}^{-}_{\lbrace a c \rbrace} \bar{m}^a \bar{m}^c .  
\end{align}
\n
The expression in the last line is obtained by decomposing the radial-frame metric perturbation $k_{AB}$ into the 2+2 gauge-invariants as per Eq. (\ref{eq.FluidFramekAB}). Immediately we can see that this expression is nothing more than the harmonic decomposition of our perturbation variable $\pazocal{J}^{-}_{ac}$ projected with $\bar{m}^a \bar{m}^c$. This is in agreement with the correspondence between the 1+1+2 master variables and the NP Weyl scalars as well as the correspondence between the 1+1+2 variables and the 2+2 gauge invariant perturbations. 

\subsubsection{Axial Perturbations and Correspondence}
In exactly the same manner as we did for the polar perturbations, we can analyse the axial sector and how the results correspond to those derived in both \cref{eqn:WeylMasterWave} and \cref{eqn:WeylMasterWaveAxial}. Luckily, the analysis is greatly simplified as the system of equations for the axial sector are less entangled than their polar sector counterparts. Perturbations to the axial sector, in vacuum, are completely characterised by a single scalar variable $\Pi$ defined by \cite{Gerlach79,Gerlach80,Gundlach00}
\begin{align}
\Pi &= \epsilon^{AB} \left( r^{-2} \, k_A \right)_{| B} \\
&= \frac{1}{r^2} \left[ - \left( 1 - \frac{2 M}{r} \right)^{-1/2} k^{\prime}_t + \frac{2}{r} k_t + \left( 1 - \frac{2 M}{r} \right)^{1/2} \dot{k}_r \right] ,
\end{align}
\n
where we have explicitly evaluated $\Pi$ in terms of its metric components for Schwarzschild coordinates. The corresponding scalar wave equation is
\begin{align}
\left[ \frac{1}{r^2} \, \left( r^4 \Pi \right)_{|A} \,  \right]^{|A} - \left( \ell + 2 \right) \left( \ell - 1 \right) &= 0 .
\end{align}
\n 
On a Cauchy surface, the two first-order degrees of freedom are given by $\lbrace \Pi , \dot{\Pi} \rbrace$. Explicitly evaluating the covariant derivatives, we would find a scalar wave equation with characteristics set by the metric: 
\begin{align}
-\ddot{\Pi} + \Pi^{\prime \prime} = \mathcal{S}_{\Pi} .
\end{align}
\n 
Consequentially, it is reasonable to state that $\Pi$ describes axial gravitational waves. Expressing the 1+1+2 variables in terms of 2+2 gauge-invariants we can immediately identify the axial master variable with the radial part of the magnetic Weyl 2-tensor
\begin{align}
\mathcal{H} &= \frac{\ell (\ell + 1)}{2} \Pi  .
\end{align}
\n
Following our previous discussion, the master variable $\pazocal{V}_{\lbrace a b \rbrace}$ is a covariant, gauge-invariant repackaging of the 2+2 gauge invariant master variable $\Pi$ whose physical significance is made manifest by the identification of $\pazocal{V}_{ab}$ with the radial part of the magnetic Weyl 2-tensor.

Finally, we can complete the correspondence by noting the link between the perturbed Weyl scalars and the 2+2 gauge invariants \cite{Nolan04}, the most significant of which is the perturbation to $\Psi_2$:
\begin{align}
\delta \Psi_2 &= - \frac{i}{4} \ell (\ell + 1) \, \Pi \, Y \\
&\sim \mathcal{H} \\ 
&\sim \Psi_{\textrm{RW}}.
\end{align}
\n
This summarises the results detailed in this paper, namely that the radial part of the magnetic Weyl Tensor corresponds to the Regge-Wheeler variable and this can be expressed in many different ways in many different approaches. To make life more difficult, it is clear from \cref{eqn:WeylMaster}) and \cref{eqn:WeylMasterAxial}) that there is a non-uniqueness in the definition of a master variable. As the scalar mode $R$ vanishes in the axial sector, these equations and correspondences will hold even in the extension to $f(R)$ gravity. 

\section{Summary}
We have presented a further analysis of linear, non-spherical perturbations to the Schwarzschild black hole in $f(R)$ gravity. Notably, we have advocated using the Weyl tensor as a powerful object for studying the evolution of gravitational waves. The two main perturbation variables that we introduce, $\pazocal{J}_{\lbrace a b \rbrace}$ and $\pazocal{V}_{\lbrace a b \rbrace}$, obey closed, covariant, gauge-invariant and frame-invariant wave equations that happen to be exactly the same in both General Relativity as well as to the $f(R)$ extensions considered here. 

The first perturbation variable $\pazocal{J}_{\lbrace a b \rbrace}$ is constructed from the transverse-traceless degrees of freedom of the Weyl tensor and is a particularly convenient way to study the evolution of gravitational perturbations in the Schwarzschild spacetime. Imposing a set of basic constraints and imposing that $f(0) = 0$ and $R=0$ in the background, the resulting equation decouples from the scalar modes at the linear level. In part this is due to the setup used, namely that we consider a vacuum Schwarzschild spacetime for which there are initially no scalar mode excitations. The scalar modes in $f(R)$ gravity correspond to massive modes, with the mass of the particles set by the parameters of the theory $f^{\prime} (0)$ and $f^{\prime \prime} (0)$, that propagate along timelike curves. The pure tensor modes described by $\pazocal{J}_{ab}$, however, will be massless and propagate along null curves. Were we to consider a more complicated spacetime, by either reducing symmetries or by introducing a matter to the system, it is likely that the scalar modes and tensor modes will be more entangled. As such, our results are in agreement with the literature on the subject \cite{DeFelice11,Myung11,Nzioki14}. The variable $\pazocal{J}_{\lbrace a b \rbrace}$ was shown to correspond to the radiative degrees of freedom in the Newman-Penrose formalism \cite{Newman62}, i.e. $\Psi_4$ and $\Psi_0$, as well as the gauge-invariant master variables $\chi$ and $\varphi$ in the 2+2 formalism \cite{Gundlach00}. This should not be of too much surprise, given that the NP Weyl scalars are projections of the Weyl tensor, which describes the free gravitational field and one of the primary reasons as to why we advocate its use in the 1+1+2 formalism. 

The second perturbation variable $\pazocal{V}_{\lbrace a b \rbrace}$ is defined to be the radial part of the magnetic Weyl tensor and is a purely axial variable. Given that the scalar modes are even parity in nature, it is unsurprising that the form for this equation is exactly the same as in General Relativity. In fact, by an appropriate rescaling, it can be shown that the radial part of the magnetic Weyl tensor is exactly the Regge-Wheeler variable. Alternatively, it can be shown that this variable is exactly $\Psi_2$ in the Newman-Penrose formalism \cite{Newman62} or $\Pi$ in the 2+2 decomposition \cite{Gundlach00}. 

Finally, we highlighted how the curvature-fluid terms can be expressed via the NP Ricci scalars, e.g. $\Phi_{00}$ and $\Phi_{11}$, which would necessarily be zero in General Relativity. 

\section{Acknowledgements}
GP would like to thank Chris Clarkson, Rituparno Goswami, Antony Lewis, Anne Marie Nzioki, David Seery and Patricia Schmidt for useful discussions and/or comments. GP would also like to thank the California Institute of Technology for hospitality where parts of this work were completed. The research leading to these results has received funding from the European Research Council under the European Union's Seventh Framework Programme (FP/2007-2013) / ERC Grant Agreement No. [616170].

\bibliography{Paper_fR.bib}

\appendix

\section{Petrov Classification}
\label{app:Petrov}
The Petrov classification allows us to invariantly divide the gravitational fields into six distinct types. The two main approaches to this problem, which are completely equivalent, are as an eigenvalue problem applied to the Weyl tensor or by looking at the principal null directions. The eigenvalue approach is tantamount to finding a set of eigenvalues $\lambda$ and eigenbivectors $X^{ab}$ such that \cite{StephaniBook}
\begin{align}
\frac{1}{2} C_{abcd} X^{cd} &= \lambda X^{ab}. 
\end{align}
\n
The eigenbivectors correspond to the principal null directions and, due to the inherent symmetries of the Weyl tensor, will belong to a four-dimensional space of antisymmetric bivectors. Using a null tetrad as per Newman-Penrose formalism, we can introduce the following set of bivectors
\begin{align}
U_{ab} &= - l_{[a} m_{b]} , \\
V_{ab} &= k_{[a} m_{b]} , \\
W_{ab} &= m_{[a} \bar{m}_{b]} - k_{[a} l_{b]} . 
\end{align}
\n
This allows us to express the Weyl tensor in terms of the NP Weyl scalars $\Psi_n$
\begin{align}
C_{abcd} &= \Psi_0 \left( U_{ab} U_{cd} \right) + \Psi_1 \left( U_{ab} W_{cd} + W_{ab} U_{cd} \right) + \Psi_2 \left( V_{ab} U_{cd} + U_{ab} V_{cd} + W_{ab} W_{cd} \right) \\
\nonumber &\qquad + \Psi_3 \left( V_{ab} W_{cd} + W_{ab} V_{cd} \right) + \Psi_4 \left( V_{ab} V_{cd} \right) .
\end{align}
\n
The Petrov classification can now be distinctly written in terms of which NP Weyl scalars vanish. The classification is as follows:
\begin{align*}
&&&{\bf{Type \, I}} &: \Psi_0 &= 0 ,& \\
&&&{\bf{Type \, II}} &: \Psi_0 &= \Psi_1 = 0 ,& \\
&&&{\bf{Type \, D}} &: \Psi_0 &= \Psi_1 = \Psi_3 = \Psi_4 = 0 ,& \\
&&&{\bf{Type \, III}} &: \Psi_0 &= \Psi_1 = \Psi_2 = 0 , &\\
&&&{\bf{Type \, N}} &: \Psi_0 &= \Psi_1 = \Psi_2 = \Psi_3 = 0 ,& \\
&&&{\bf{Type \, O}} &: \Psi_0 &= \Psi_1 = \Psi_2 = \Psi_3 = \Psi_4 = 0 . &
\end{align*}
\n
In terms of the Petrov type, the Weyl peeling theorem can be re-expressed as follows
\begin{align}
C_{abcd} &\sim \frac{N}{r} + \frac{\textrm{Type III}}{r^2} + \frac{\textrm{Type II}}{r^3} + \frac{\textrm{Type I}}{r^4} + \mathcal{O} \left( r^{-5} \right) .
\end{align}
\n
It should be noted, however, that the peeling behaviour only holds for spacetimes that are weakly asymptotically simple and can be extended smoothly to $\mathscr{J}^{\pm}$, though this may be too strict in certain cases \cite{Kroon98}. 

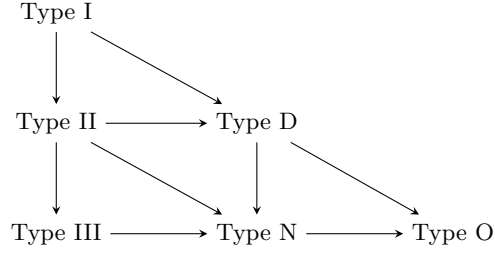
\begin{figure}
\begin{tikzpicture}
  \matrix (m) [matrix of math nodes,row sep=3em,column sep=4em,minimum width=2em]
  {
     \textrm{Type I} & \phantom{TEXT} & \phantom{TEXT}  \\
     \textrm{Type II} & \textrm{Type D} & \phantom{TEXT}  \\
     \textrm{Type III} & \textrm{Type N} & \textrm{Type O} \\};
  \path[-stealth]
    (m-2-1.east|-m-2-2) edge node [below] {} node [above] {} (m-2-2)
    (m-1-1) edge node [above,right] {$\; \; $} (m-2-2) 
    (m-2-2) edge node [above,right] {$\; \; $} (m-3-3) 
    (m-2-2) edge node [below] {} (m-3-2) 
    (m-1-1) edge node [below] {} (m-2-1)
    (m-2-1) edge node [below] {} (m-3-2) 
    (m-2-1) edge node [below] {} (m-3-1)
    (m-3-2) edge node [below] {$$} (m-3-3) 
    (m-3-1.east|-m-3-2) edge node [below] {} node [above] {} (m-3-2)
    ;
\end{tikzpicture}
\caption{Conventional diagram showing the degenerations of the Petrov type \cite{StewartBook,StephaniBook}.}
\end{figure}

\section{Radiated Energy}
\label{app:Energy}
The correspondence between the electric and magnetic Weyl 2-tensors and the NP Weyl scalar $\Psi_4$ allows us to look at the normal equations for the radiated energy flux and radiated momenta. For a 2-surface at infinity $S^2$, the radiated energy flux is related to $\Psi_4$ as follows
\begin{align}
\frac{d E}{dt} &= \lim_{r \rightarrow \infty} \left\lbrace \frac{r^2}{4 \pi} \int_{S^2} \, d \Omega \, \left| \int^t_{-\infty} \Psi_4 \, d t^{\prime} \right|^2 \right\rbrace \\
&= \lim_{r \rightarrow \infty} \left\lbrace \frac{r^2}{4 \pi} \int_{S^2} \, d \Omega \, \left| \int^t_{-\infty} \pazocal{J}^{-}_{\lbrace a b \rbrace} \bar{m}^a \bar{m}^b \, d t^{\prime} \right|^2 \right\rbrace .
\end{align}
\n
Likewise, the radiated linear momenta can be expressed via a radial unit vector $\tilde{r}_i = (\sin \theta \cos \phi , \sin \theta \sin \phi, \cos \theta)$ in flat space as follows
\begin{align}
\frac{d P_i}{dt} &= \lim_{r \rightarrow \infty} \left\lbrace \frac{r^2}{16 \pi} \int_{S^2} \, d \Omega \, \tilde{r}_i \, \left| \int^t_{-\infty} \Psi_4 \, d t^{\prime} \right|^2 \right\rbrace \\
&= \lim_{r \rightarrow \infty} \left\lbrace \frac{r^2}{16 \pi} \int_{S^2} \, d \Omega \, \tilde{r}_i \, \left| \int^t_{-\infty} \pazocal{J}^{-}_{\lbrace a b \rbrace} \bar{m}^a \bar{m}^b \, d t^{\prime} \right|^2 \right\rbrace .
\end{align}

\section{Harmonics}
\label{app:harmonics}
\subsection{Scalar Harmonics:}
A set of dimensionless covariant harmonics $Q$ can be defined on any LRS background as the eigenfunctions of the 2-dimensional Laplace-Beltrami operator \cite{Clarkson03,Clarkson07}:
\begin{align}
\delta^2 Q &= - \frac{k^2}{r^2} Q , \qquad \hat{Q} = \dot{Q} = 0 \qquad ( k^2 \geq 0 ) .
\end{align}
\n
The radial parameter $r$ can be covariantly defined by
\begin{align}
\hat{r} &= \frac{r}{2} \phi , \qquad \dot{r} = -\frac{r}{2} \left( \Sigma - \frac{2}{3} \Theta \right) , \qquad \delta_a r = 0 .
\end{align}
\n
Any first order scalar may now be decomposed into this scalar harmonic basis 
\begin{align}
\psi &= \displaystyle\sum_k \psi^{(k)}_S Q^{(k)} = \psi_S Q .
\end{align}

\subsection{Vector Harmonics:}
Even parity vector harmonics can be systematically constructed from the scalar harmonics 
\begin{align}
Q^{(k)}_a &= r \, \delta_a Q^{(k)} \Rightarrow \hat{Q}_a = \dot{Q}_a = 0 , \\
\delta^2 Q_a &= \frac{1}{r^2} \left( 1 - k^2 \right) Q_a ,
\end{align}
\n
and odd parity vector harmonics are defined by
\begin{align}
\bar{Q}^{(k)}_a &= r \, \epsilon_{ab} \delta^b Q^{(k)} \Rightarrow \hat{\bar{Q}}_a = \dot{\bar{Q}}_a = 0 , \\
\delta^2 \bar{Q}_a &= \frac{1}{r^2} \left( 1 - k^2 \right) \bar{Q}_a .
\end{align}
\n
By construction, these harmonics are orthogonal parity inversions of each other, $\bar{Q}_a = \epsilon_{ab} Q^b \Leftrightarrow Q_a = - \epsilon_{ab} \bar{Q}^b$ and $Q^a \bar{Q}_a = 0$. A crucial difference arises in that the even parity harmonics are not solenoidal but the odd parity harmonics are:
\begin{align}
\delta^a Q_a &= - \frac{k^2}{r} Q , \qquad \\
\delta^a \bar{Q}_a &= 0 , \\
\epsilon_{ab} \delta^a Q^b &= 0, \\
\epsilon_{ab} \delta^a \bar{Q}^b &= \frac{k^2}{r} Q . 
\end{align}
\n
Any first order vector may be decomposed into this basis
\begin{align}
\psi_a &= \displaystyle\sum_k \psi^{(k)}_S Q^{(k)}_a + \bar{\psi}_V \bar{Q}^{(k)}_a = \psi_V Q_a + \bar{\psi}_V \bar{Q}_a .
\end{align}

\subsection{Tensor Harmonics:}
Likewise, we can extend these definitions to construct even and odd parity tensor harmonics
\begin{align}
Q_{ab} &= r^2 \delta_{\lbrace a} \delta_{b \rbrace} Q \, \Rightarrow \, \hat{Q}_{ab} = \dot{Q}_{ab} = 0, \\
\bar{Q}_{ab} &= r^2 \epsilon_{c \lbrace a} \delta^c \delta_{b \rbrace} Q \, \Rightarrow \hat{\bar{Q}}_{ab} = \dot{\bar{Q}}_{ab} = 0 .
\end{align}
\n
As with the vector harmonics, these tensor harmonics will be orthogonal parity inversions of one another: $Q_{ab} = - \epsilon_{c \lbrace a} \bar{Q}_{b \rbrace}^{\phantom{b \rbrace}c} \Leftrightarrow \bar{Q}_{ab} = \epsilon_{c \lbrace a} Q_{b \rbrace}^{\phantom{b \rbrace} c}$ and $Q^{ab} \bar{Q}_{ab} = 0$. Any first order tensor may be harmonically expanded in this basis
\begin{align}
\psi_{ab} &= \displaystyle\sum_k \psi_T^{(k)} Q^{(k)}_{ab} + \bar{\psi}_T^{(k)} \bar{Q}^{(k)}_{ab} = \psi_T Q_{ab} + \bar{\psi}_T \bar{Q}_{ab} .
\end{align}

\subsection{Relation to Spherical Harmonics}
The scalar covariant harmonics are defined on a sphere of radius $r$, whereas the conventional spherical harmonics are defined on the unit sphere. This allows us to construct a correspondence between the covariant and spherical harmonics.

\phantom{TEXT}
\n
For scalar harmonics:
\begin{align}
Q &= Y .
\end{align}
For vector harmonics:
\begin{align}
Q_a &= r Y_a , \\
\bar{Q}_a &= r \bar{Y}_a .
\end{align}
For tensor harmonics:
\begin{align}
Q_{ab} &= r^2 Y_{ab} , \\
\bar{Q}_{ab} &= - \frac{1}{2} r^2 \bar{Y}_{ab} ,
\end{align}
\n
where the last line depends on the exact definition of the axial tensorial harmonics (i.e. sometimes the prefactor of $-1/2$ is omitted). 
\end{document}